\documentclass{ws-ijbc}

\usepackage{graphicx}
\usepackage{dcolumn}
\usepackage{bm}
\newcommand{\be}{\begin{equation}}
\newcommand{\ee}{\end{equation}}
\usepackage{dsfont}
\usepackage{color}
\newcounter{commentzaehler}
\DeclareMathOperator{\const}{const}

\renewcommand\epsilon{\varepsilon}
\renewcommand\le{\leqslant}
\renewcommand\ge{\geqslant}
\newcommand{\trn}{^{\rm\scriptscriptstyle T}}
\newcommand{\e}{\rm e}

\begin{document}
\catchline{}{}{}{}{} 

\markboth{S.~A.~Plotnikov et al.}{Adaptive control of synchronization
  in delay-coupled  heterogeneous networks of FitzHugh-Nagumo nodes }

\title{Adaptive control of synchronization in delay-coupled heterogeneous networks of FitzHugh-Nagumo nodes }
\author{S.~A.~Plotnikov}
\address{Department of Theoretical Cybernetics, Saint-Petersburg State University, St.~Petersburg, Russia}
\author{J.~Lehnert}
\address{Institut f\"{u}r Theoretische Physik, TU Berlin,
Hardenbergstra{\ss}e 36, D-10623 Berlin, Germany}
\author{A.~L.~Fradkov}
\address{Department of Theoretical Cybernetics, Saint-Petersburg State University, St.~Petersburg, Russia}
\address{ Institute for Problems
of Mechanical Engineering, Russian Academy of Sciences,
Bolshoy~Ave,~61, Vasilievsky Ostrov,~St.~Petersburg, 199178~Russia }
\author{E.~Sch\"oll}
\address{Institut f\"{u}r Theoretische Physik, TU Berlin,
Hardenbergstra{\ss}e 36, D-10623 Berlin, Germany}
\date{\today}
\maketitle
\begin{abstract}
We study synchronization in delay-coupled neural networks
of heterogeneous nodes. It is well known that heterogeneities in the nodes
hinder synchronization when becoming too large.  We show that an adaptive tuning of the overall
coupling strength can be used to counteract the effect of the
heterogeneity. Our adaptive controller is demonstrated  on ring networks
of FitzHugh-Nagumo systems which are  paradigmatic for  excitable dynamics
but can also -- depending on the system parameters -- exhibit self-sustained periodic firing.  We show that the adaptively tuned time-delayed
coupling enables synchronization even if parameter heterogeneities are  so large
that excitable nodes coexist with  oscillatory ones. 
\end{abstract}


\section{Introduction}

The ability to control nonlinear dynamical systems has brought up a
wide interdisciplinary area of research that has evolved rapidly in
the past decades \cite{SCH07}. Besides the control of isolated
systems, control of dynamics in spatiotemporal systems and on networks
has recently gained much interest
\cite{SUN13,FLU10b,OME11,POS13a,BAC14}.  Adaptive control schemes have
emerged as a new type of control methods that enable control in
situations where parameters are unknown or drift in time. They allow
for automatically tuning the parameters to appropriate values and are,
therefore, of particular interest for experiments and technological
applications.  Previously, they have been successfully applied in the
control of network dynamics\cite{SEL12,LEH14}. Here, we show  that adaptive
control methods can be
used to counteract the effect of heterogeneous nodes in the
synchronization of delay-coupled networks.

Synchronization in neural networks has gained a lot of attention
lately \cite{PIK03} since it is involved in processes as diverse as
learning and visual perception on the one hand
\cite{FRI05a,UHL09,SIN99} and the occurrence of Parkinson's disease
and epilepsy on the other hand \cite{TAS98,POE01,UHL09}. Control of
synchronization has so far focused on networks of identical nodes
\cite{ZHO08,LU09, LU12, SEL12, GUZ13, LEH14}. Further results on the
control of network dynamics  were obtained in \cite{LU15,LI14}. 
However, in realistic
networks the nodes will always be characterized by some diversity
meaning that the parameters of the different nodes are not identical
but drawn from a distribution.  It is well known that such
heterogeneities in the nodes can hinder or prevent synchronization and
that the coupling strength is a crucial parameter in this context
\cite{STR00,SUN09a}. Moreover, in most cases perfect synchronization -- in the sense
that the state of all nodes is identical at all times  -- is unfeasible in
the presence of  heterogeneities.
Although quite a number of papers are devoted to synchronization 
of heterogeneous networks, see Refs.~\cite{DEL15,ISI14, CHE14,RIC12} and references therein, only a few of them address 
adaptive control of synchronization  \cite{GUO13, FRA13a}. Moreover, the adaptation of the coupling strength was not considered
previously in the case of heterogeneous networks.
Here, we develop an adaptive controller which
allows for reaching a synchronization state where the node's
trajectories are close though not identical.

Our method is based on the speed-gradient (SG) method, which was
previously used in the control of delay-coupled networks
\cite{SEL12,GUZ13,LEH14}, however, not in the presence of node
heterogeneities. 
In order to apply the SG method, we suggest a goal
function which characterizes the quality of synchronization. 
Based on this measure an adaptive controller is
developed which ensures synchronization even if the parameter
heterogeneities become such large that some nodes -- if uncoupled --
undergo a Hopf bifurcation and behave 
distinctly different from the other nodes in the network.  We
demonstrate our algorithm on the FitzHugh-Nagumo (FHN) system
\cite{FIT61a,NAG62}, a generic model for neural dynamics.
The advantage of our approach is that the SG method allows for
using a simple goal function which does not depend upon the parameters
to be controlled but only on the system state. 

Note that in contrast to artificial neural networks Refs.~\cite{LI14,LU15,SHI15} which are designed to solve advanced
computational tasks, we here assume that the network, i.e., the local
dynamics and the network topology, are given and only design the
algorithm adapting the coupling strength. Furthermore,  usually all nodes in artificial neural
networks are identical, while we try to mimic
realistic networks and therefore use heterogeneous network nodes.

The paper is organized as follows: Section \ref{sec:speed-grad-meth} is a recapitulation of the
SG method, while Sec.~\ref{sec:model-equation} introduces the
model. Section~\ref{sec:two-delay-coupled} discusses two
delay-coupled FHN systems: The possible bifurcation
scenarios are investigated and the adaptive control algorithm is
developed.  In Sec.~\ref{sec:adapt-synchr-ring}, the method is generalized
to larger ring networks. Finally, we conclude with Sec.~\ref{sec:conclusion}.

\section{Speed-Gradient Method}\label{sec:speed-grad-meth}
In this section, we briefly review the speed-gradient (SG) method \cite{FRA07}. Consider a general nonlinear dynamical system
\begin{equation}\label{f1}
\dot{\mathbf{x}}=\mathbf{F}(\mathbf{x},\mathbf{g},t)
\end{equation}
with state vector $\mathbf{x}\in \mathbb{C}^n$, input (control) variables $\mathbf{g}
\in \mathbb{C}^m$, and nonlinear function $\mathbf{F}$. Define a control goal
\begin{equation}\label{f2}
Q(\mathbf{x}(t),t)\le \Delta,
\end{equation}
for $t\ge t^*$, where $Q(\mathbf{x},t)\ge0$ is a smooth scalar goal function and $\Delta$ is
the desired level of precision. For example, if we want to force the
trajectory of system \eqref{f1} to follow the desired trajectory
$\mathbf{x}^*(t)$, we can use a goal function in the form $Q(\mathbf{x}(t))=(\mathbf{x}(t)-\mathbf{x}^*(t))^2$.

In order to design a control algorithm, the scalar function
$\dot{Q}=\mathbf{\omega}(\mathbf{x},\mathbf{g},t)$ is calculated, that is, the speed (rate) at
which $Q(\mathbf{x}(t),t)$ is changing along the trajectories of Eq.~\eqref{f1}:
\begin{equation}\label{f3}
\omega(\mathbf{x},\mathbf{g},t)=\frac{\partial Q(\mathbf{x},t)}{\partial t}+[\nabla_\mathbf{x}Q(\mathbf{x},t)]\trn \mathbf{F}(\mathbf{x},\mathbf{g},t).
\end{equation}
Then the gradient of $\omega(\mathbf{x},\mathbf{g},t)$ with respect to
the input variables
is evaluated as 
\begin{equation}\label{f4}
\nabla_\mathbf{g}\omega(\mathbf{x},\mathbf{g},t)=\nabla_\mathbf{g}[\nabla_\mathbf{x}Q(\mathbf{x},t)]\trn \mathbf{F}(\mathbf{x},\mathbf{g},t).
\end{equation}
Finally, we obtain the control function $\mathbf{g}$ from
\begin{equation}\label{f5}
\mathbf{g}(t)=\mathbf{g}^0-\psi(\mathbf{x},\mathbf{g},t),
\end{equation}
where the vector function $\psi(\mathbf{x},\mathbf{g},t)=\gamma \nabla_\mathbf{g}\omega(\mathbf{x},\mathbf{g},t)$ with some adaptation gain $\gamma >0$, and
$\mathbf{g}^0=\const$ is an initial (reference) control value (often
$\mathbf{g}^0=0$ is assumed). The algorithm \eqref{f5} is called
\textit{speed-gradient} (SG) \textit{algorithm in finite form} since
it suggests to change $\mathbf{g}$ proportionally to the gradient of the speed
of changing $Q$. For the \textit{speed-gradient} (SG) in its
\textit{differential form} see Ref.~\cite{FRA07}.

Several analytic conditions exist guaranteeing that the control goal
\eqref{f2} can be achieved in system \eqref{f1} and \eqref{f5}. The main condition is the existence of a constant value of the parameter $\mathbf{g}^*$, ensuring attainability of the goal in the system $d\mathbf{x}/dt=\mathbf{F}(\mathbf{x},\mathbf{g}^*,t)$. Details can be found in the control-related literature \cite{FRA79,SHI00a}.

The idea of this algorithm is the following: The term
$-\nabla_\mathbf{g}\omega(\mathbf{x},\mathbf{g},t)$ points to the
direction in which the value of $\dot{Q}$ decreases with the highest
speed. Therefore, if one forces the control signal to "follow" this
direction, the value of $\dot{Q}$ will decrease and finally be
negative. When $\dot{Q}<0$, then $Q$ will decrease and, eventually,
will tend to zero.

\section{Model equation}\label{sec:model-equation}
The local dynamics of each node in the network is modeled by the
FitzHugh-Nagumo (FHN) differential equations \cite{FIT61a,NAG62}. The FHN model is paradigmatic for excitable dynamics close to a Hopf bifurcation \cite{LIN04}, which is not only characteristic for neurons but also occurs in the context of other systems ranging from electronic circuits \cite{HEI10} to cardiovascular tissues and the climate system \cite{MUR93,IZH00a}. Each node of the network is described as follows:
\begin{equation}\label{m}
\begin{aligned}
\epsilon \dot{u_i}&=u_i-\frac{u_i^3}{3}-v_i+C\sum\limits_{j=1}^{N}G_{ij}[u_j(t-\tau)-u_i(t)], \\
\dot{v_i}&=u_i+a_i,\quad i=1,\dots,N,
\end{aligned}
\end{equation}
where $u_i$ and $v_i$ denote the activator and inhibitor variable of
the nodes $i=1,\dots,N$, respectively. $\tau$ is the delay, i.e., the
time the signal needs to propagate between node  $i$ and $j$  (here we
will use $\tau=1.5$). $\epsilon$ is a time-scale parameter and
typically small (here we will use $\epsilon=0.1$), i.e., $u_i$
is a fast variable, while $v_i$ changes slowly. The coupling matrix $\mathbf{G}=\{G_{ij}\}$
defines which nodes are connected to each other. We construct the
matrix $\mathbf{G}$ by setting the entry $G_{ij}$ equal to $1$~(or $0$) if
the $j$th node couples (or does not couple) into the $i$th node. After repeating
this procedure for all entries  $G_{ij}$, we normalize each
row to unity. The overall coupling strength is given by
$C$.

In the uncoupled system ($C=0$), $a_i$ is a threshold parameter: For
$a_i>1$ the $i$th node of the system is excitable, while for $a_i<1$
it exhibits self-sustained periodic firing. This is due to a
supercritical Hopf bifurcation at $a_i=1$ with a locally stable equilibrium
point for $a_i>1$ and a stable limit cycle for $a_i<1$. In previous
publications, networks of homogeneous FHN systems were considered,
i.e., $a_1=a_2=\ldots=a_N\equiv a$   \cite{BRA09, SCH08, HOE09,
  LEH11, PAN12,CAK14}. In particular, it was shown that for excitable systems, i.e., $a>1$
and coupling matrices with positive entries zero-lag synchronization
is always a stable solution independently of the coupling strength and
delay time (as long as both are large enough to induce any spiking at
all).

 Here, we investigate the case of heterogeneous nodes. In this case,
 perfect synchronization, i.e., $ (u_1,v_1)= \ldots = (u_N,v_N)
 \equiv (u_s,v_s)$, is no longer a solution of Eq.~\eqref{m}
 which can  easily be  seen by  plugging $ (u_1,v_1)= \ldots = (u_N,v_N)
 \equiv (u_s,v_s)$ into Eq.~\eqref{m}. The node dynamics is then
 described by
\begin{equation}\label{ms}
\begin{aligned}
\epsilon \dot{u_s}&=u_s-\frac{u_s^3}{3}-v_s+C \sum [u_s(t-\tau)-u_s(t)], \\
\dot{v_s}&=u_s+a_i,\quad i=1,\dots,N,
\end{aligned}
\end{equation}
which is obviously not independent of $i$. This means that a perfectly synchronous
solution does not exist in system ~\eqref{m} because the prerequisite for the
existence of such a solution is that each node receives the same input
if all nodes are in synchrony.  However, solutions close
to the synchronous solution might exist where the nodes spike at the same
(or almost the same) time but with slightly different amplitudes. As we show, these
solutions can be reached and stabilized by an adaptive tuning of the coupling strength.

\section{Two delay-coupled FitzHugh-Nagumo systems}\label{sec:two-delay-coupled}
This Section studies the most basic network motif consisting of two
coupled systems without self-feedback.
Before deriving the adaptive controller, we perform a linear stability
analysis of the equilibrium point to get insight in the possible bifurcations.

\subsection{Linear stability of the equilibrium point} \label{sec:line-stab-fixed}

The linear stability analysis follows the approach suggested in
Ref.~\cite{DAH08c,SCH08}. 
Consider two coupled FHN-systems with heterogeneous
threshold parameters and bidirectional coupling
\begin{equation}\label{m1}
\begin{aligned}
\epsilon \dot{u_1}&=u_1-\frac{u_1^3}{3}-v_1+C[u_2(t-\tau)-u_1(t)], \\
\dot{v_1}&=u_1+a_1, \\
\epsilon \dot{u_2}&=u_2-\frac{u_2^3}{3}-v_2+C[u_1(t-\tau)-u_2(t)], \\
\dot{v_2}&=u_2+a_2.
\end{aligned}
\end{equation}

The unique equilibrium point of the system \eqref{m1} is given by
$\mathbf{x}^*\equiv(u_1^*,v_1^*,u_2^*,v_2^*)^T$, where $u_1^*=-a_1$,
$u_2^*=-a_2$, $v_1^*=-a_1+a_1^3/3+C(a_1-a_2)$ and
$v_2^*=-a_2+a_2^3/3+C(a_2-a_1)$.
Linearizing Eq.~\eqref{m1} around the equilibrium point $\mathbf{x}^*$
 by setting $\mathbf{x}(t)=[u_1(t),v_1(t),u_2(t),v_2(t)]^T\equiv
\mathbf{x}^*+\delta\mathbf{x}(t)$, we obtain
\begin{equation}\label{m2}
\delta \mathbf{\dot x}=\frac{1}{\epsilon}\begin{pmatrix}
\xi_1 & -1 & 0 & 0 \\
\epsilon & 0 & 0 & 0 \\        
0 & 0 & \xi_2 & -1 \\
0 & 0 & \epsilon & 0
\end{pmatrix}\delta\mathbf{x}(t)+\frac{1}{\epsilon}\begin{pmatrix}
0 & 0 & C & 0 \\
0 & 0 & 0 & 0 \\        
C & 0 & 0 & 0 \\
0 & 0 & 0 & 0
\end{pmatrix}\delta\mathbf{x}(t-\tau),
\end{equation}
where $\xi_i=1-a_i^2-C$. The ansatz
\begin{equation}\label{m3}
\delta \mathbf{x}(t)=\e^{\lambda t}\mathbf{q},
\end{equation}
where $\mathbf{q}$ is a time-independent vector, leads to the characteristic equation for the eigenvalues $\lambda$,
\begin{equation}\label{m4}
(1-\xi_1\lambda+\epsilon\lambda^2)(1-\xi_2\lambda+\epsilon\lambda^2)-(\lambda C\e^{-\lambda\tau})^2=0.
\end{equation}

A necessary condition for the occurrence of a Hopf
bifurcation is that the eigenvalue $\lambda$ is imaginary. 
We, therefore, substitute  the ansatz $\lambda=i \omega$, $\omega \in
\mathds{R}$, into Eq.~\eqref{m4} to find the threshold parameters $a_1$ and
$a_2$ for which a Hopf bifurcation can take place.  Then, separating  Eq.~\eqref{m4}
into real and imaginary parts yields 
\begin{equation}\label{m5}
\begin{aligned}
(1-\epsilon\omega^2)^2-\xi_1\xi_2\omega^2&=-\omega^2C^2\cos(2\omega\tau),\\
\omega(1-\epsilon\omega^2)(\xi_1+\xi_2)&=-\omega^2C^2\sin (2\omega\tau).
\end{aligned}
\end{equation}

Since $\epsilon\ll1$, terms of the order of $\epsilon$ can neglected, i.e., we consider $\epsilon=0$ in the
following.  Squaring and adding the subequations of Eq.~\eqref{m5} we obtain
\begin{equation}\label{m6}
\omega^2(\xi_1+\xi_2)^2+(1-\xi_1\xi_2\omega^2)^2=\omega^4C^4,
\end{equation}
which can be rearranged to
\begin{equation}\label{m7}
(\xi_1^2\xi_2^2-C^4)\omega^4+(\xi_1^2+\xi_2^2)\omega^2+1=0.
\end{equation}

Equation \eqref{m7} is biquadratic, i.e. quadratic in $z \equiv \omega^2$. Therefore, we can use Vieta's
formulas to analyze whether  it has  non-negative roots. According to
Vieta the following holds
\begin{equation}\label{m8}
z_1+z_2=-\frac{\xi_1^2+\xi_2^2}{\xi_1^2\xi_2^2-C^4}, \qquad z_1z_2=\frac{1}{\xi_1^2\xi_2^2-C^4},
\end{equation}
where $z_1$ and $z_2$ are roots of the quadratic equation \eqref{m7}. For
$\xi_1^2\xi_2^2>C^4$,  $z_1<0$ and $z_2<0$ follows meaning that
Eq.~\eqref{m7} has no real-valued solution $\omega$, and,
thus, no Hopf bifurcation will take place. Taking the square root of this inequality and resubstituting $\xi_i=1-a_i^2-C$ yields
\begin{equation}\label{m9}
|(1-C-a_1^2)(1-C-a_2^2)|>C^2.
\end{equation}

\begin{figure}[ht]
\flushleft
\begin{minipage}[h]{0.24\linewidth}
\center{\includegraphics[width=1\linewidth]{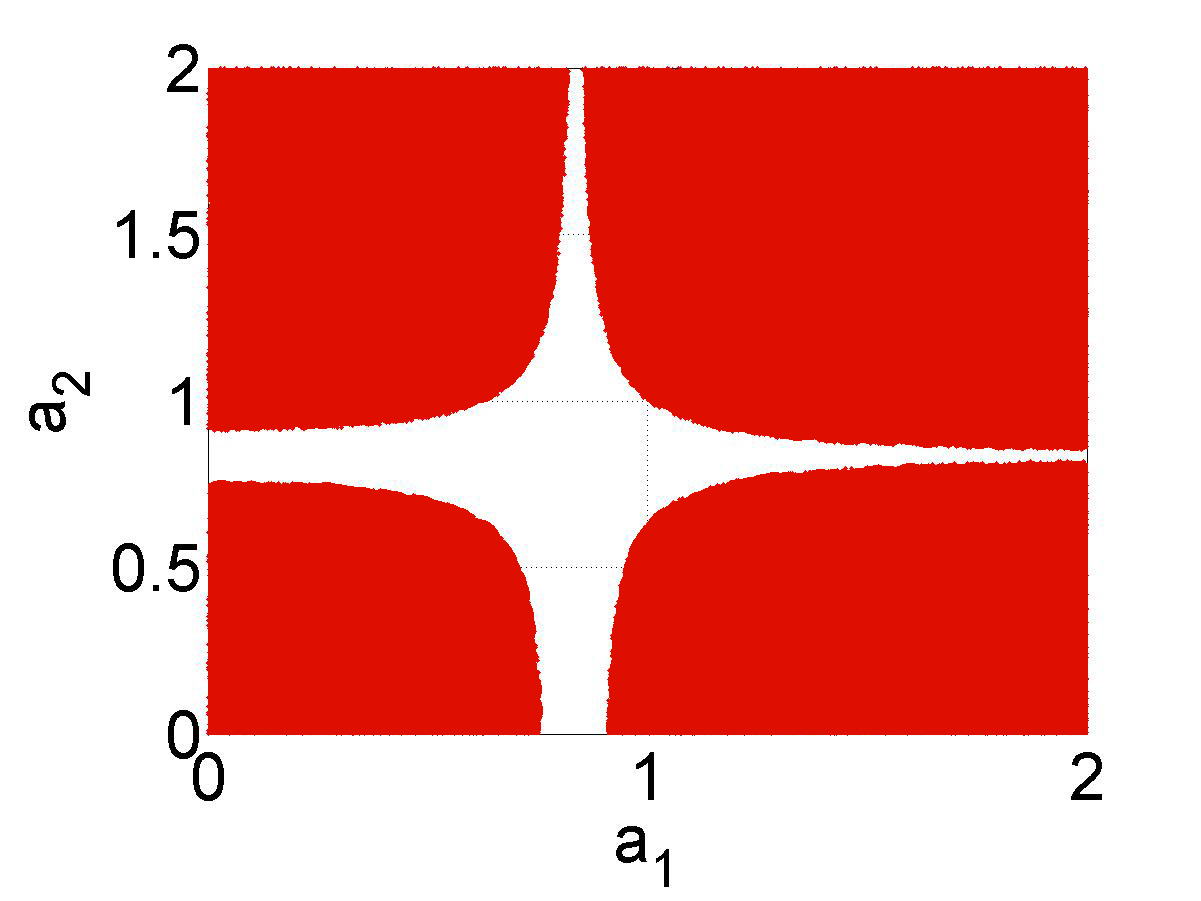} \\ (a)}
\end{minipage}
\begin{minipage}[h]{0.24\linewidth}
\center{\includegraphics[width=1\linewidth]{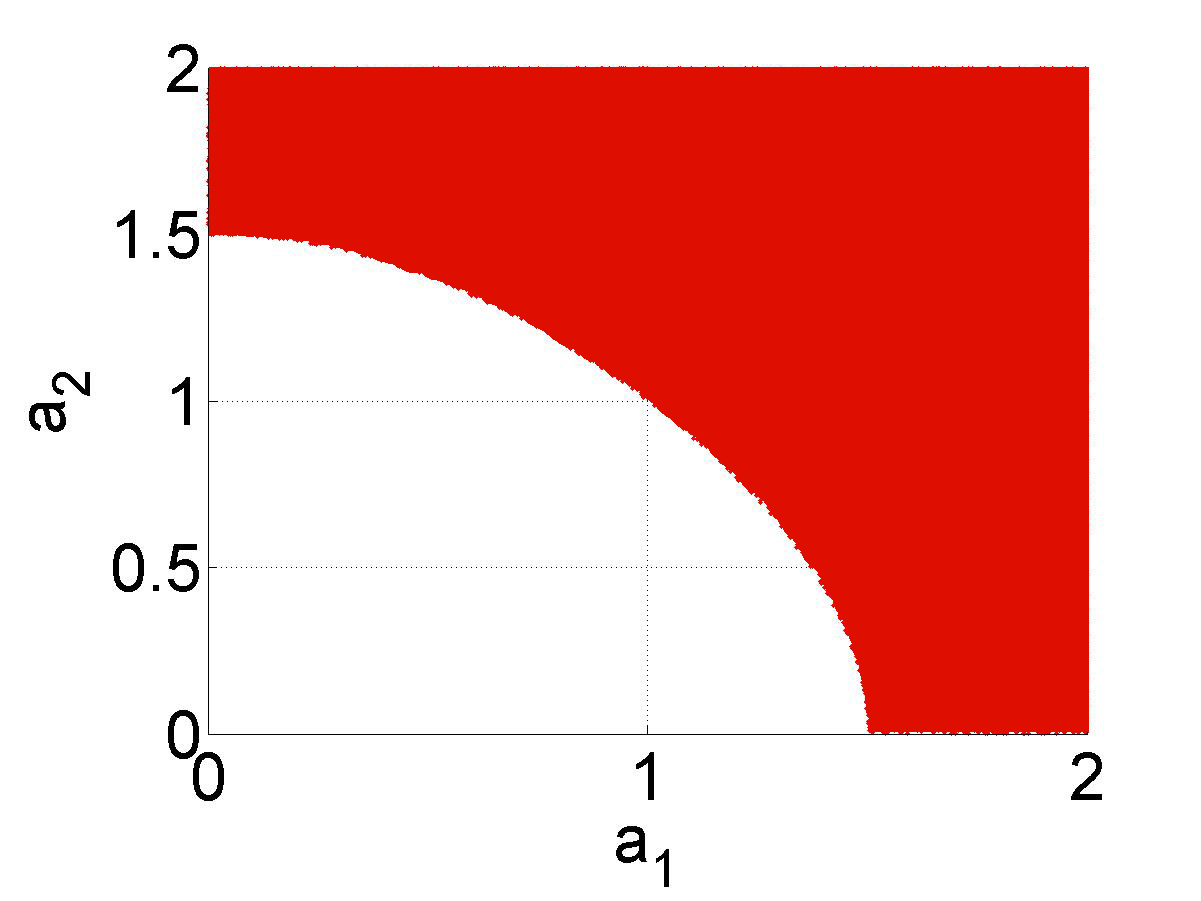} \\ (b)}
\end{minipage}
\caption{Hopf bifurcation of the equilibrium point of two delay-coupled FHN
  systems (Eq.~\eqref{m1}) for (a)  small coupling strength $C$, i.e., $C=0.3$, and (b) large coupling
  strength $C=5$. (Red) shading marks parameter values for which inequality
  \eqref{m9} is fulfilled, i.e., a Hopf bifurcation is impossible
  and the equilibrium point is stable. 
}\label{figh}
\end{figure}

This inequality defines the values of $a_1$ and $a_2$ where a
Hopf bifurcation is impossible, i.e., the equilibrium point is stable. Figure
\ref{figh} shows this area in (red) shading for (a) small coupling strength,
i.e., $C<1$, and (b) large coupling strength, i.e.,  $C>1$. As
expected no Hopf bifurcation will occur for $a_1>1$ and $a_2>1$ since
then both systems are in the excitable regime. Note that, however,
oscillations can coexist with the stable equilibrium point \cite{LEH11}. 



\subsection{Adaptive control of two coupled FHN-systems}\label{sec:adaptive-control-two}
We now want to apply the SG method to system \eqref{m1} with the goal
to synchronize  the two heterogeneous nodes.  As discussed  above
perfect synchronization in the form $(u_1,v_1)=(u_2, v_2)$ is not
attainable in this case but the two systems will follow slightly
different trajectories in the synchronized case. We, therefore, use as
a goal function 
\begin{equation}\label{ff9}
Q(\mathbf{x}(t),t)=\frac{1}{2}(u_1(t)-u_2(t)+a_1-a_2)^2.
\end{equation}
The choice \eqref{ff9}  ensures that the
system follows trajectories for which
\begin{subequations}
\begin{eqnarray}\label{ff8}
u_1(t)-u_2(t)&\approx& -a_1+a_2, \\
v_1(t)-v_2(t) &\approx& c\label{ff8b}
\end{eqnarray}
\end{subequations}
holds for $t\ge t^*$, where $c$ is a  constant. 
Approximation \eqref{ff8}  directly follows from the chosen goal
function \eqref{ff9}; approximation \eqref{ff8b}  is obtained by
plugging \eqref{ff8} into Eq.~\eqref{m1}. Thus, the goal
function~\eqref{ff9} yields synchronization with a shift in
the values of the inhibitors and activators of the two nodes.

From Eq.~\eqref{f5} with $\mathbf{g}=C$, system~\eqref{m1}, goal function~\eqref{ff9}, and $\psi(\mathbf{x},C,t)=\gamma \nabla_C\omega(\mathbf{x},C,t)$
 an adaptive law is straightforwardly derived:
\begin{eqnarray}\label{f10}
C(t) &=& C_0 +\frac{\gamma}{\epsilon}\left[u_1(t)-u_2(t)+a_1-a_2\right]
\left[u_1(t)-u_2(t)+u_1(t-\tau)-u_2(t-\tau)\right]
\end{eqnarray}
where $\gamma>0$ is the gain and $C_0$ is the initial value of the control parameter.  The appropriate value of $\gamma$ has to be determined by numerical simulations.
Note that a similar approach has been used to tune the coupling strength in a
network of R\"ossler systems in Ref.~\cite{GUZ13}. 

Close to the control goal,  $u_1(t)-u_2(t)\sim
u_1(t-\tau)-u_2(t-\tau)\sim a_1-a_2$ holds. We, therefore, can simplify the
adaptation law by substituting the delayed variables by their
non-delayed versions  and obtain
\begin{equation}\label{f11}
C(t) = C_0+\frac{2\gamma}{\epsilon}\left[u_1(t)-u_2(t)+a_1-a_2\right]\left[u_1(t)-u_2(t)\right].
\end{equation}


\begin{figure}
\flushleft
\begin{minipage}[h]{0.24\linewidth}
\center{\includegraphics[width=1\linewidth]{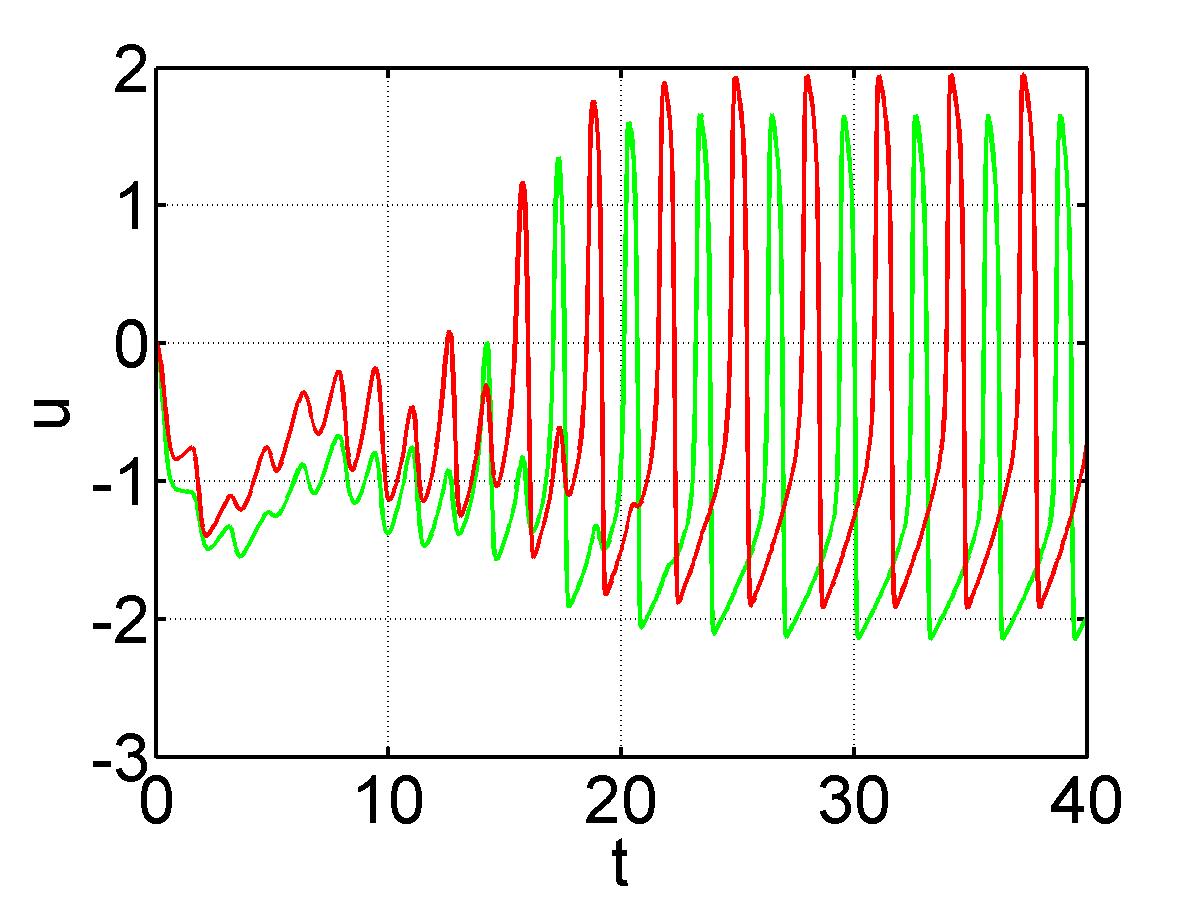} \\ (a)}
\end{minipage}
\begin{minipage}[h]{0.24\linewidth}
\center{\includegraphics[width=1\linewidth]{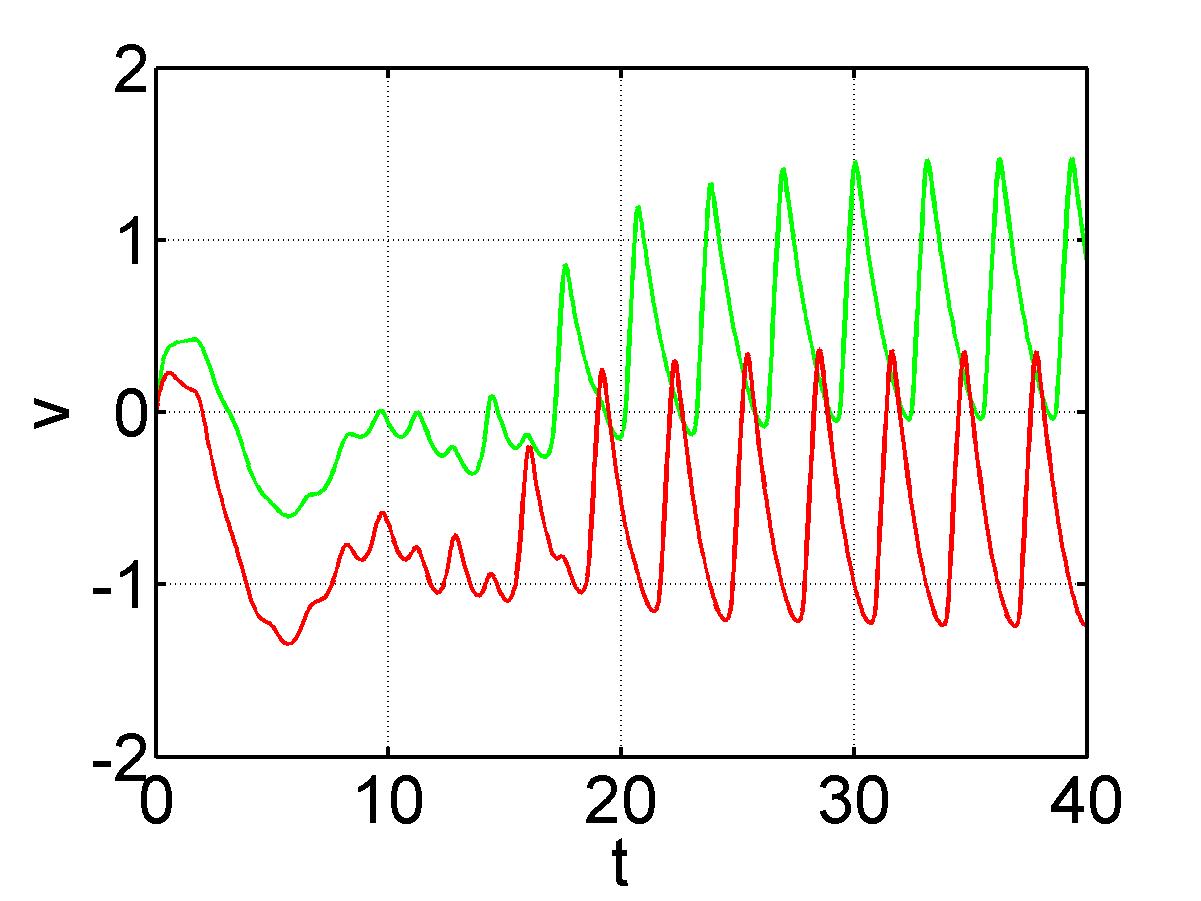} \\ (b)}
\end{minipage}
\vfill
\begin{minipage}[h]{0.24\linewidth}
\center{\includegraphics[width=1\linewidth]{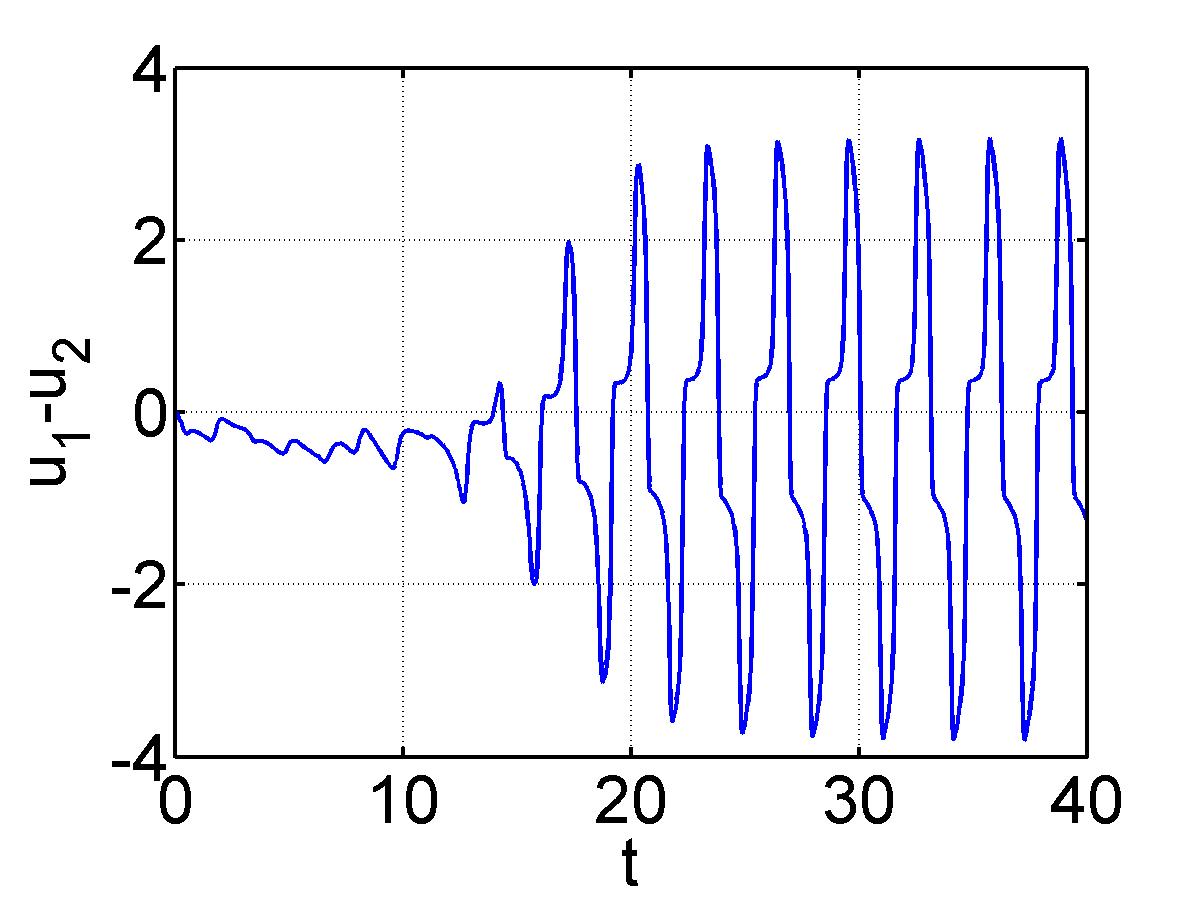} \\ (c)}
\end{minipage}
\begin{minipage}[h]{0.24\linewidth}
\center{\includegraphics[width=1\linewidth]{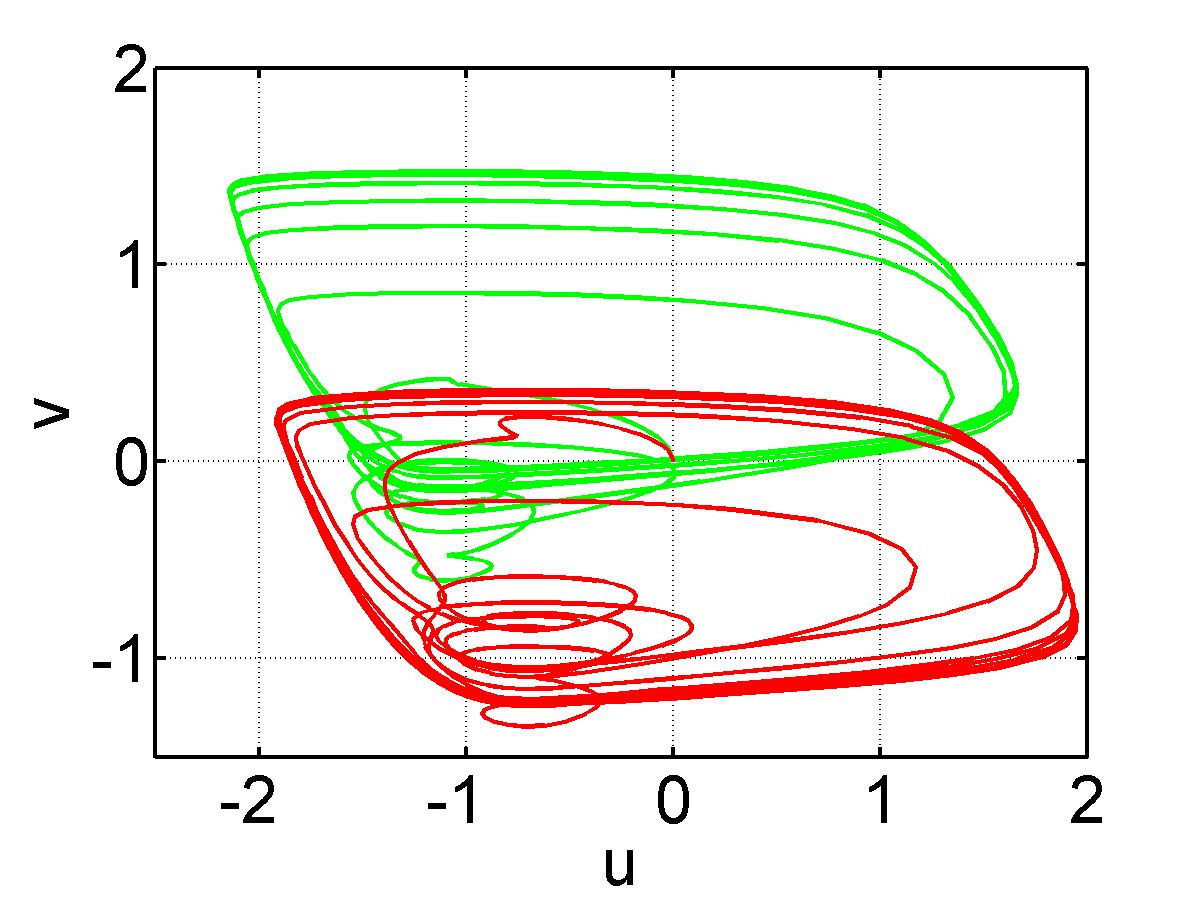} \\ (d)}
\end{minipage}
\caption{Dynamics of two coupled FitzHugh-Nagumo systems according to
  Eq.~\eqref{m1} with constant coupling strength $C$. Gray (green) 
  line marks node one, black (red)  line  marks node two. (a) and (b): time series of the activator
  and the inhibitor, respectively; (c): difference $u_1-u_2$ between the activator values, and (d): phase space. Parameters:
  $N=2$, $\epsilon=0.1$, $\tau=1.5$, $a_1=1.1$, $a_2=0.7$,
  $C=1$. Initial conditions: $u_i(t)=v_i(t)=0$, $i=1,2$, for $t \in [-\tau,0]$.}
\label{fig1}
\end{figure} 

\begin{figure}
\flushleft
\begin{minipage}[h]{0.24\linewidth}
\center{\includegraphics[width=1\linewidth]{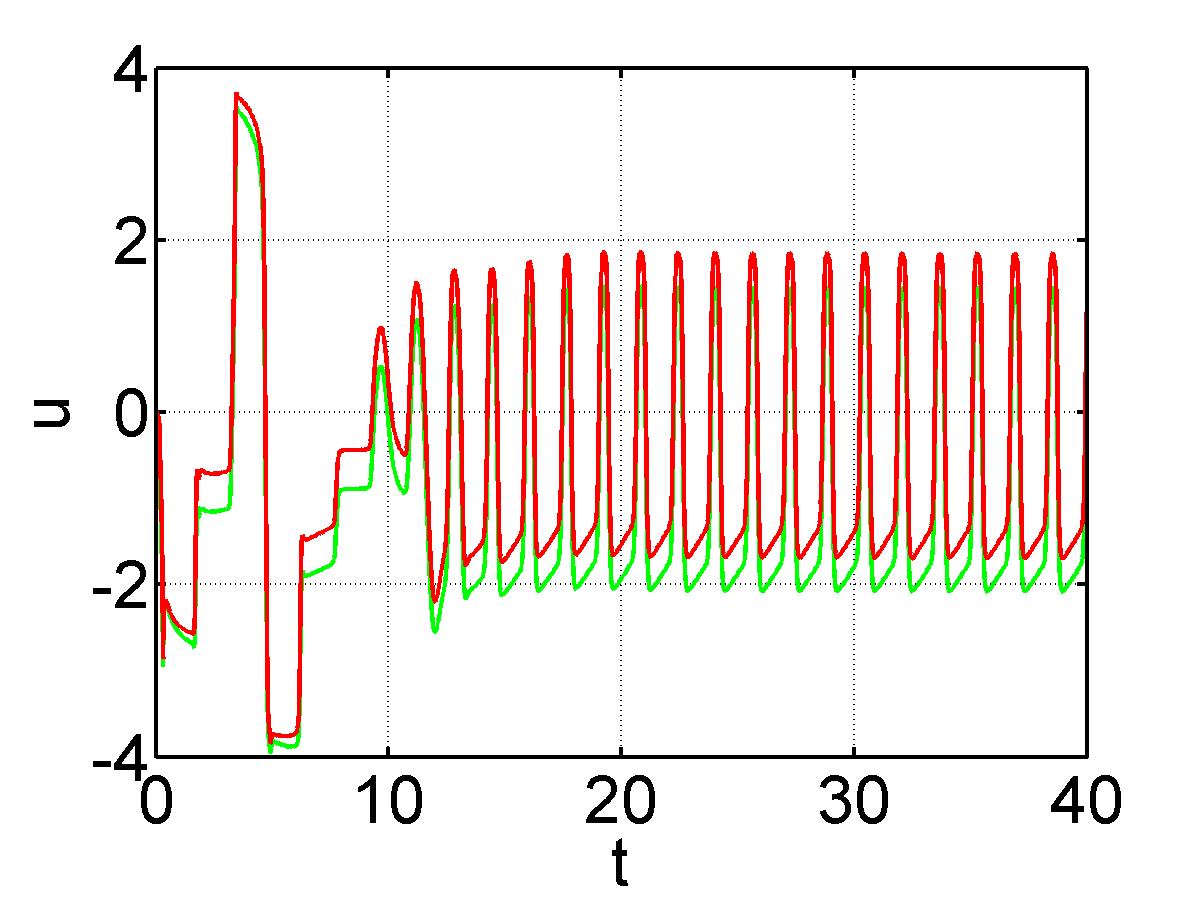} \\ (a)}
\end{minipage}
\begin{minipage}[h]{0.24\linewidth}
\center{\includegraphics[width=1\linewidth]{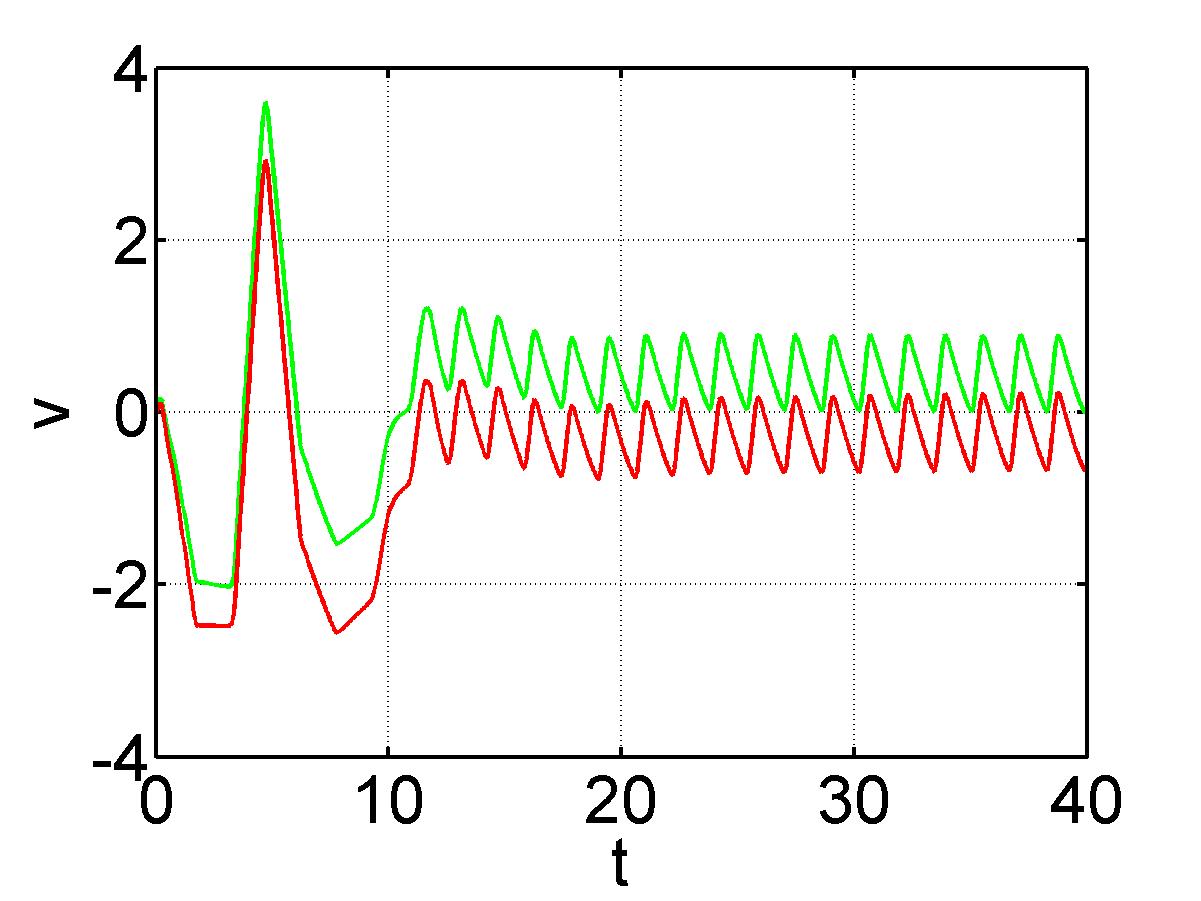} \\ (b)}
\end{minipage}
\vfill
\begin{minipage}[h]{0.24\linewidth}
\center{\includegraphics[width=1\linewidth]{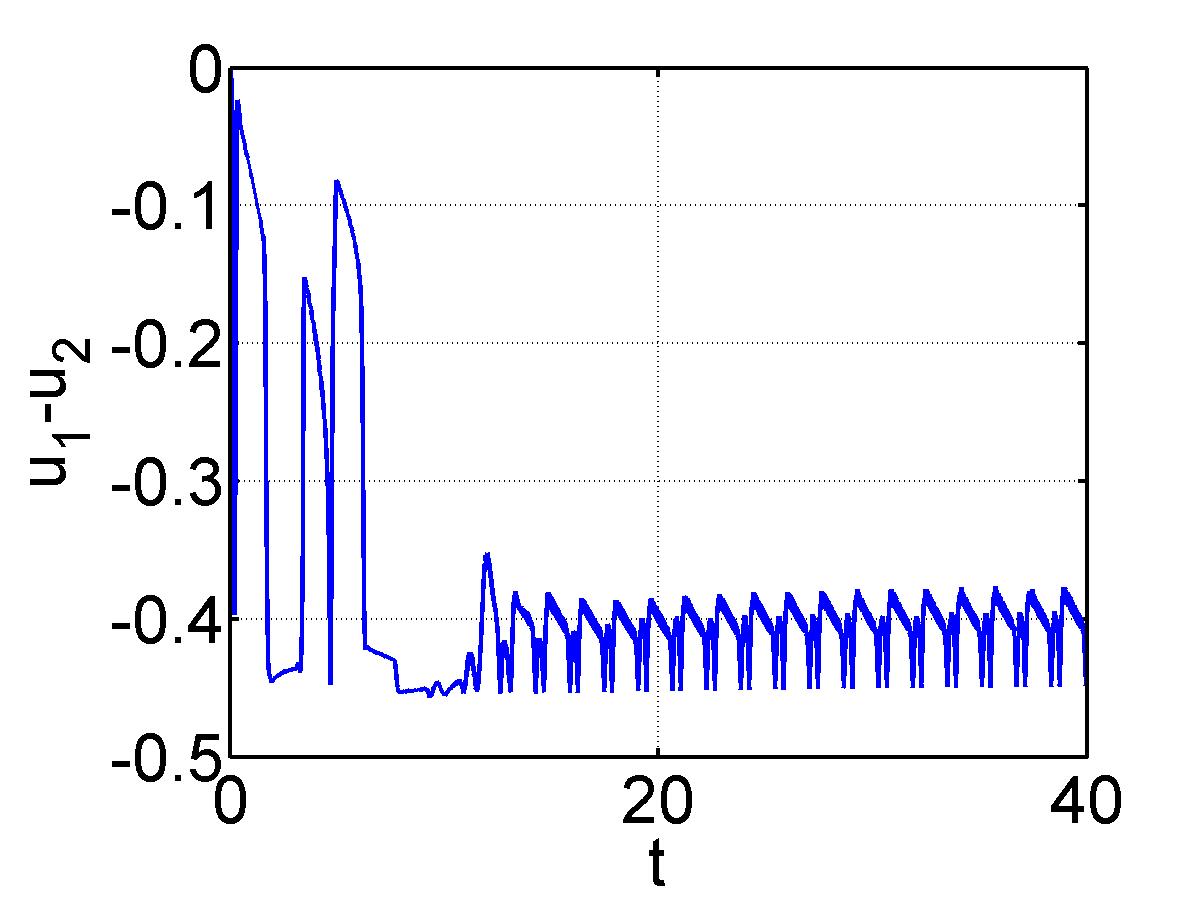} \\ (c)}
\end{minipage}
\begin{minipage}[h]{0.24\linewidth}
\center{\includegraphics[width=1\linewidth]{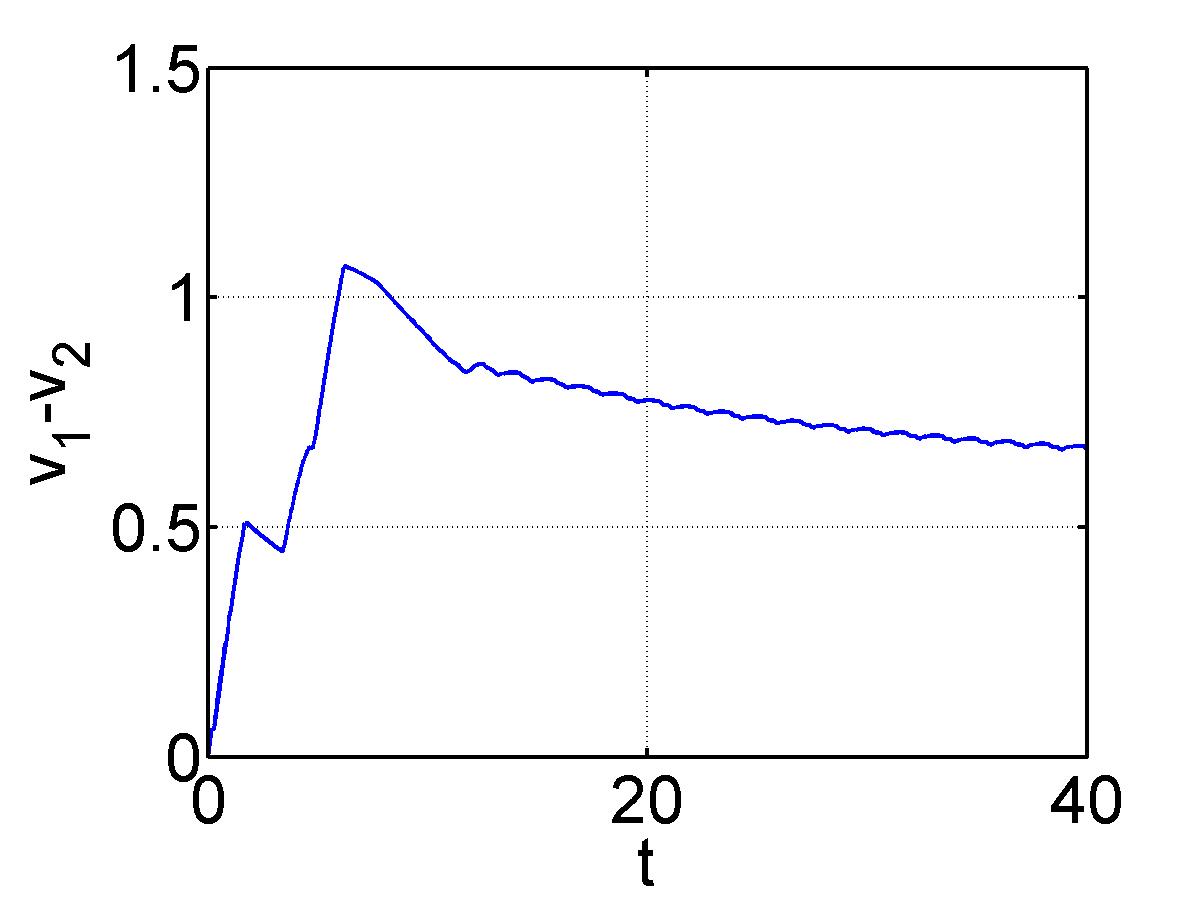} \\ (d)}
\end{minipage}
\vfill
\begin{minipage}[h]{0.24\linewidth}
\center{\includegraphics[width=1\linewidth]{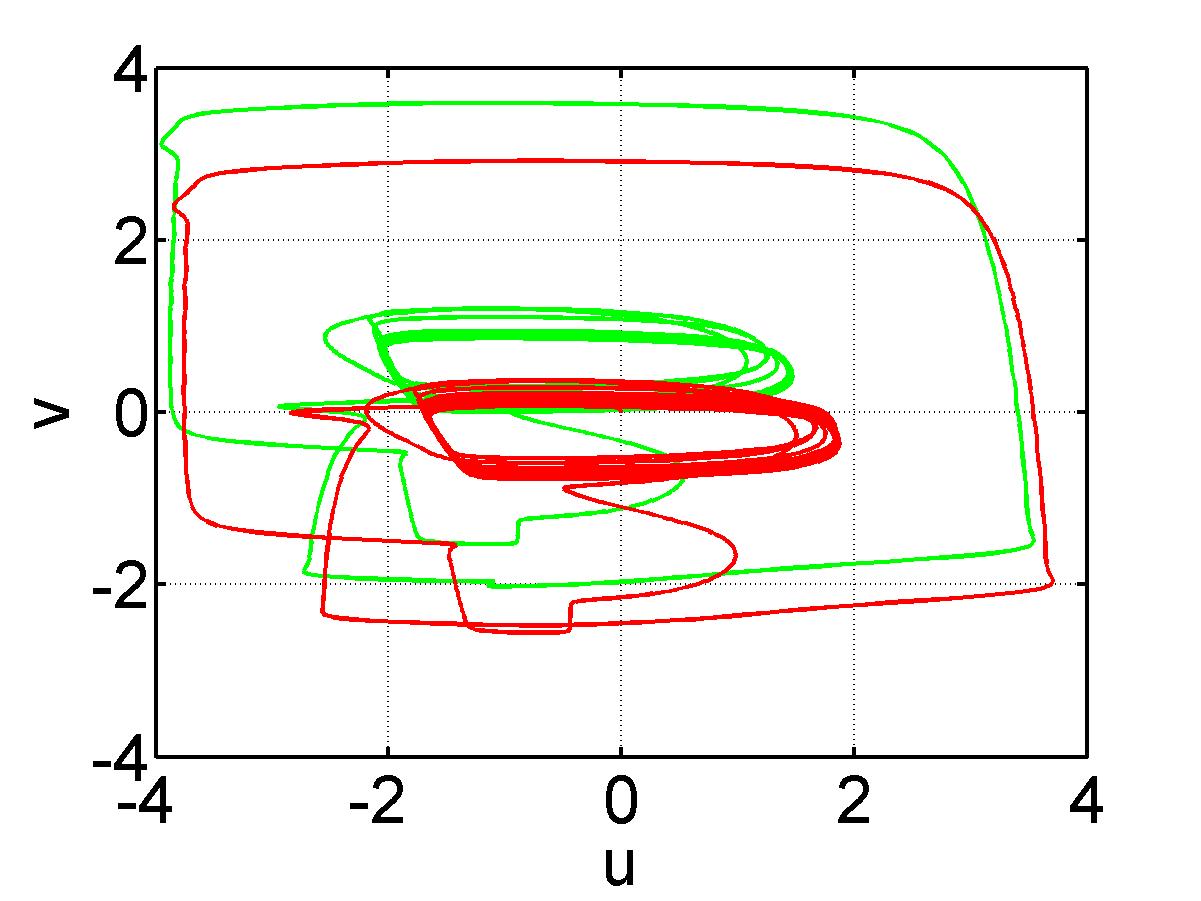} \\ (e)}
\end{minipage}
\begin{minipage}[h]{0.24\linewidth}
\center{\includegraphics[width=1\linewidth]{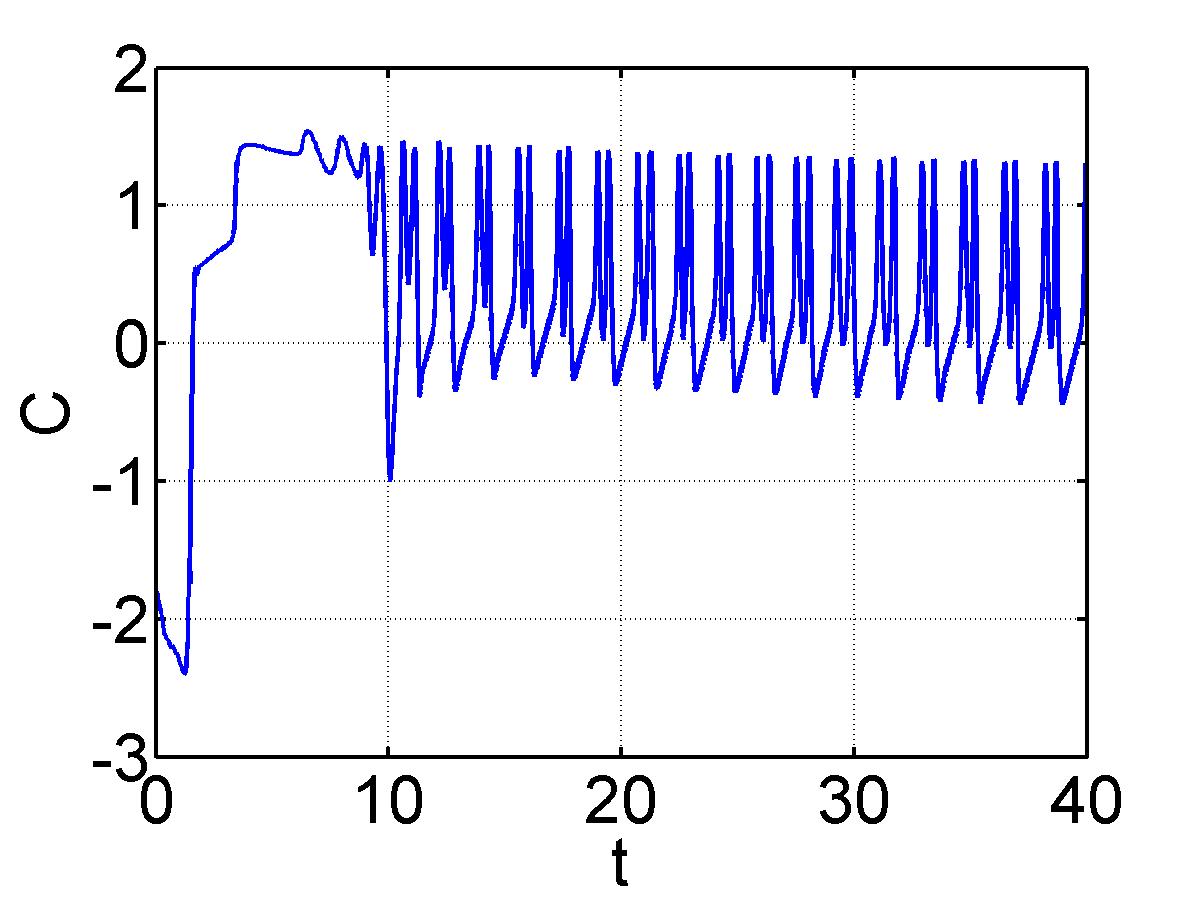} \\ (f)}
\end{minipage}
\caption{Adaptive control of two coupled FitzHugh-Nagumo systems
  (Eq.~\eqref{m1}). Gray (green) 
  line marks node one, black (red)  line marks node two. 
(a) and (b): time series of the activator
  and the inhibitor, respectively; (c) and (d): differences $u_1-u_2$ and $v_1-v_2$ between the activator and the inhibitor values, respectively; (e): phase space, and (f): time series of the coupling strength adapted according to Eq.~\eqref{f11}. Parameters: $\gamma=3$, $C_0=0$. Other
parameters and initial conditions as in Fig~\ref{fig1}.}
\label{fig2}
\end{figure}

\begin{figure}
\begin{minipage}[h]{0.24\linewidth}
\center{\includegraphics[width=1\linewidth]{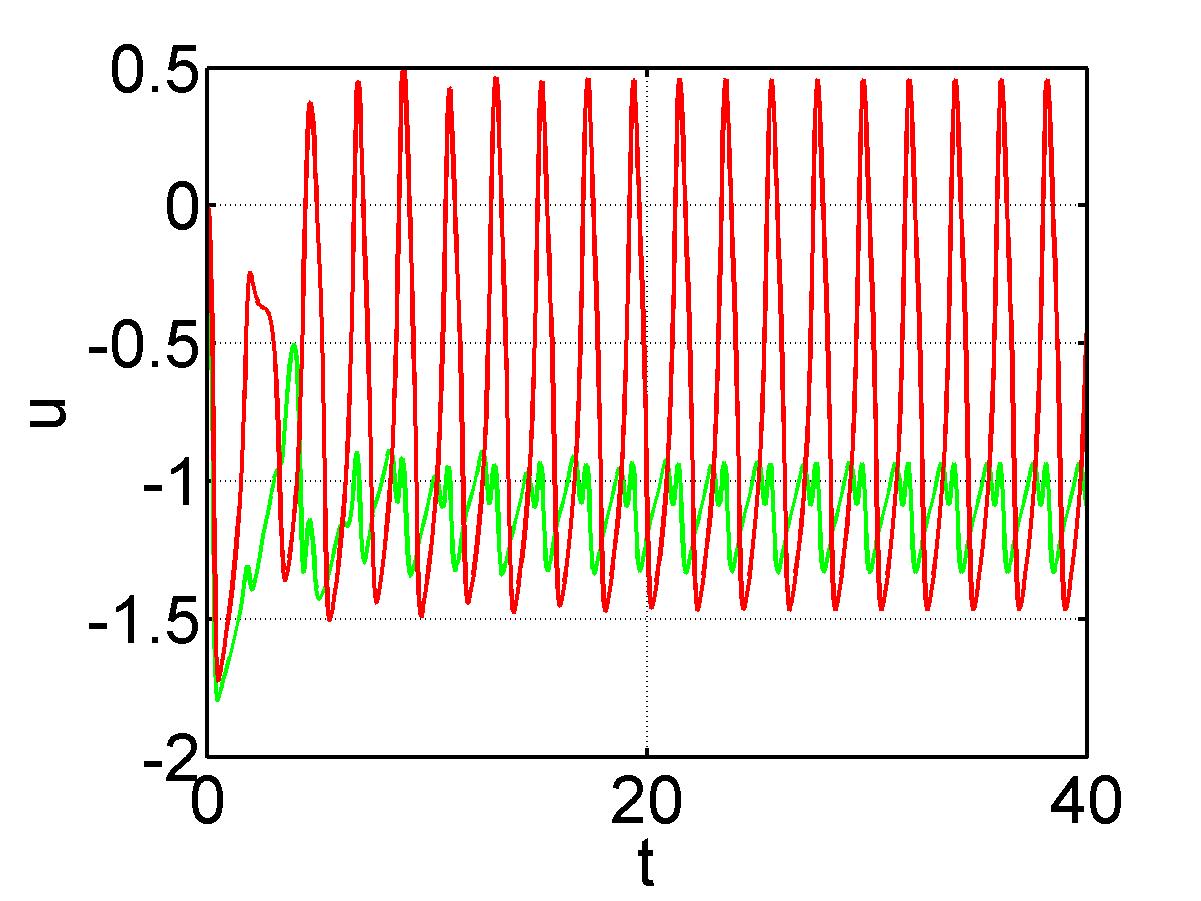} \\ (a)}
\end{minipage}
\begin{minipage}[h]{0.24\linewidth}
\center{\includegraphics[width=1\linewidth]{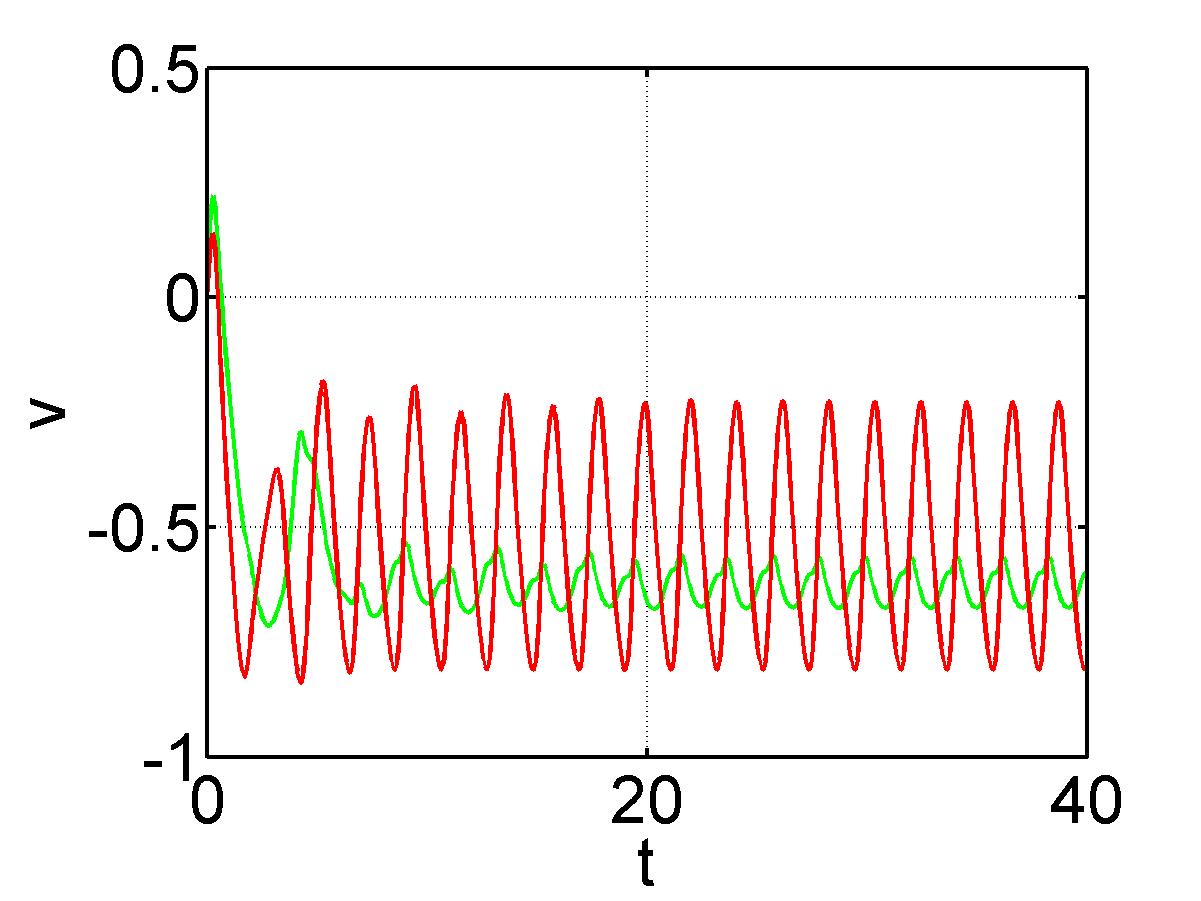} \\ (b)}
\end{minipage}
\vfill
\begin{minipage}[h]{0.24\linewidth}
\center{\includegraphics[width=1\linewidth]{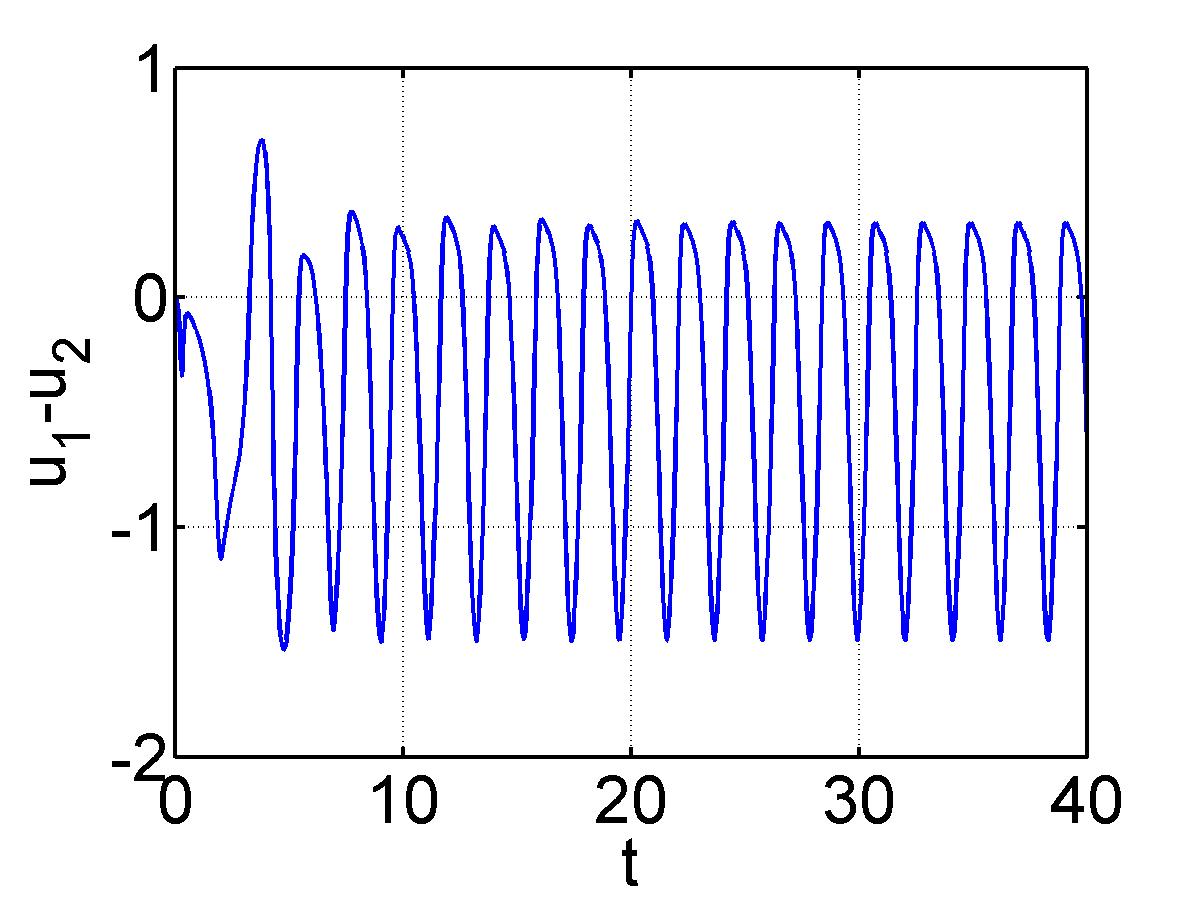} \\ (c)}
\end{minipage}
\begin{minipage}[h]{0.24\linewidth}
\center{\includegraphics[width=1\linewidth]{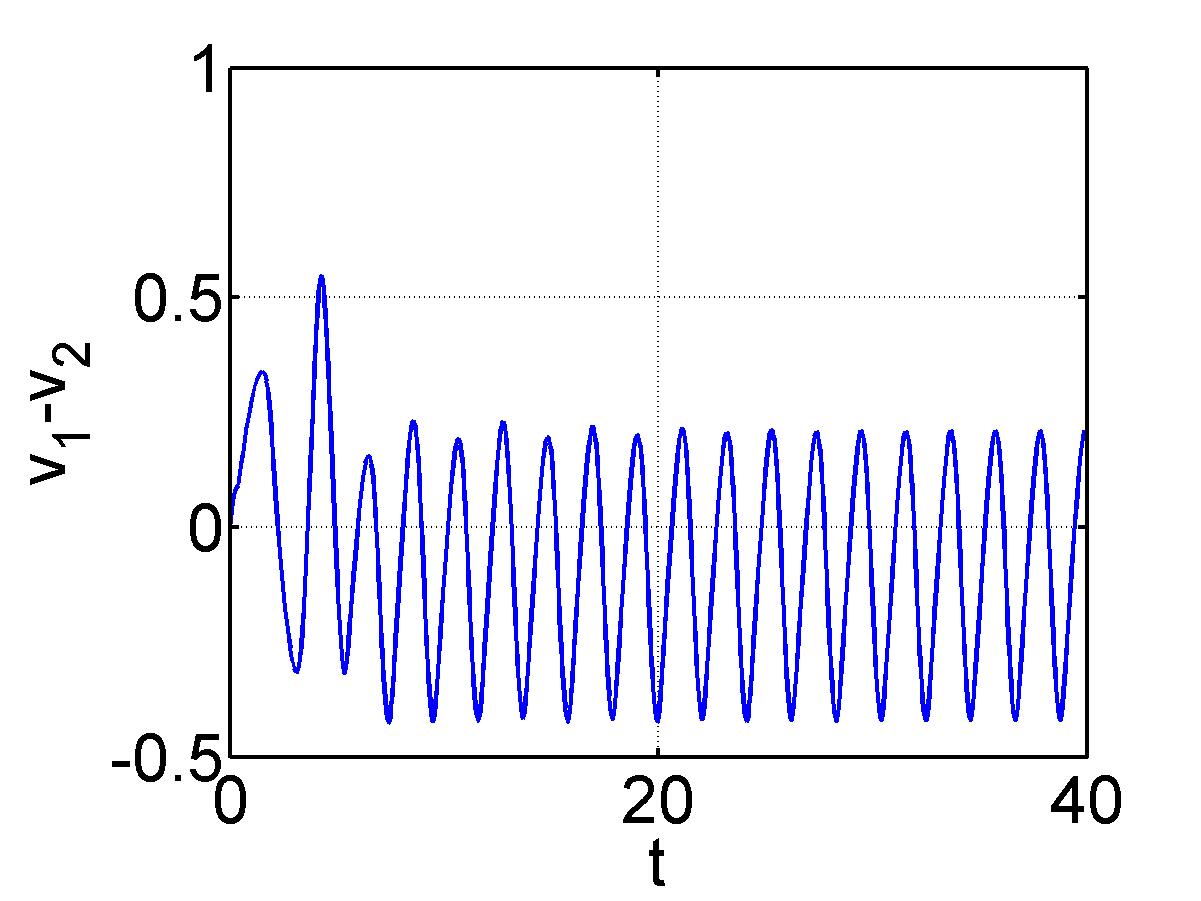} \\ (d)}
\end{minipage}
\caption{As in Fig.~\ref{fig2} but with smaller adaptation gain
  $\gamma=0.025$. 
 Gray (green)
  line marks node one, black (red)  line marks node two. 
(a) and (b): time series of the activator
  and the inhibitor, respectively; (c) and (d): differences $u_1-u_2$
  and $v_1-v_2$ between the activator and the inhibitor values, respectively.}
\label{figad}
\end{figure}

For constant coupling strength, i.e., $\gamma=0$, the two  coupled
FHN systems do not synchronize in-phase, but approach an anti-phase synchronized
state:  Figure~\ref{fig1} shows  in panels (a) and (b) the  time series of
the activators and the inhibitors, respectively, and  in panel  (d) the phase
portrait. Though the first
node is in the excitable regime ($a_1=1.1>1$) both nodes oscillate
due to the nonzero coupling strength $C$. However, they  do not
synchronize as can clearly be seen in panel (c) which depicts the
difference $u_1-u_2$ between the activator values. Instead, they phase
lock with a phase
shift of approximately $\pi$ which corresponds to an anti-phase synchronized state.

We now adapt the coupling strength  according to Eq.~\eqref{f11} in
order to synchronize the two systems
where the result is shown in Fig.~\ref{fig2}. After a transient time of
approximately 15 units of time the two systems reach the desired
synchronized state  (see the time series of the activator in
Fig.~\ref{fig2}(a) and the difference between its values Fig.~\ref{fig2}(c)). Thus, the
control is successful. 

If the gain $\gamma$ is chosen too low the control fails:
Figure~\ref{figad} depicts the results of the adaptive control according to
Eq.~\eqref{f11} for
$\gamma=0.05$. Clearly, the control does not succeed in  synchronizing the two
systems (see the time series of the activators in  Fig.~\ref{figad}(a)
and the difference between their values in (c)).

The adaptive controller~\eqref{f10} ensures  synchronization of the
activators with a shift given by $a_2-a_1$ (see
Eq.~\eqref{ff8}). Furthermore,  there is a finite, constant shift in
the inhibitor values (see Eq.~\eqref{ff8b}). The shift in the inhibitor
values can be reduced to a value close to zero if we  control the
coupling strength of each node separately. The two FHN systems are then described by
\begin{equation}\label{ad1}
\begin{aligned}
\epsilon \dot{u_i}&=u_i-\frac{u_i^3}{3}-v_i+C_i(t)[u_{(i+1) \bmod 2}(t-\tau)-u_i(t)], \\
\dot{v_i}&=u_i+a_i,\quad i=1,2
\end{aligned}
\end{equation}
where $C_i(t)$ describes the strength of the coupling to node~$i$. From Eq.~\eqref{f5} with $\mathbf{g}=(C_1,C_2)$, system~\eqref{ad1}, goal function~\eqref{ff9}, and $\psi(\mathbf{x},\mathbf{g},t)=\gamma \nabla_\mathbf{g}\omega(\mathbf{x},\mathbf{g},t)$
 an adaptive law is straightforwardly derived:
\begin{multline}\label{ad2}
C_i(t) = C_i^0 +\frac{\gamma}{\epsilon}\left[u_i(t)-u_{(i+1) \bmod 2}(t)+a_i-a_{(i+1) \bmod 2}\right]
\left[ u_i(t)-u_{(i+1) \bmod 2}(t-\tau)\right],\quad i=1,2
\end{multline}
where $\gamma>0$ is the gain and $C_i^0$ is the initial value of
control parameter. 
\begin{figure}
\begin{minipage}[h]{0.24\linewidth}
\center{\includegraphics[width=1\linewidth]{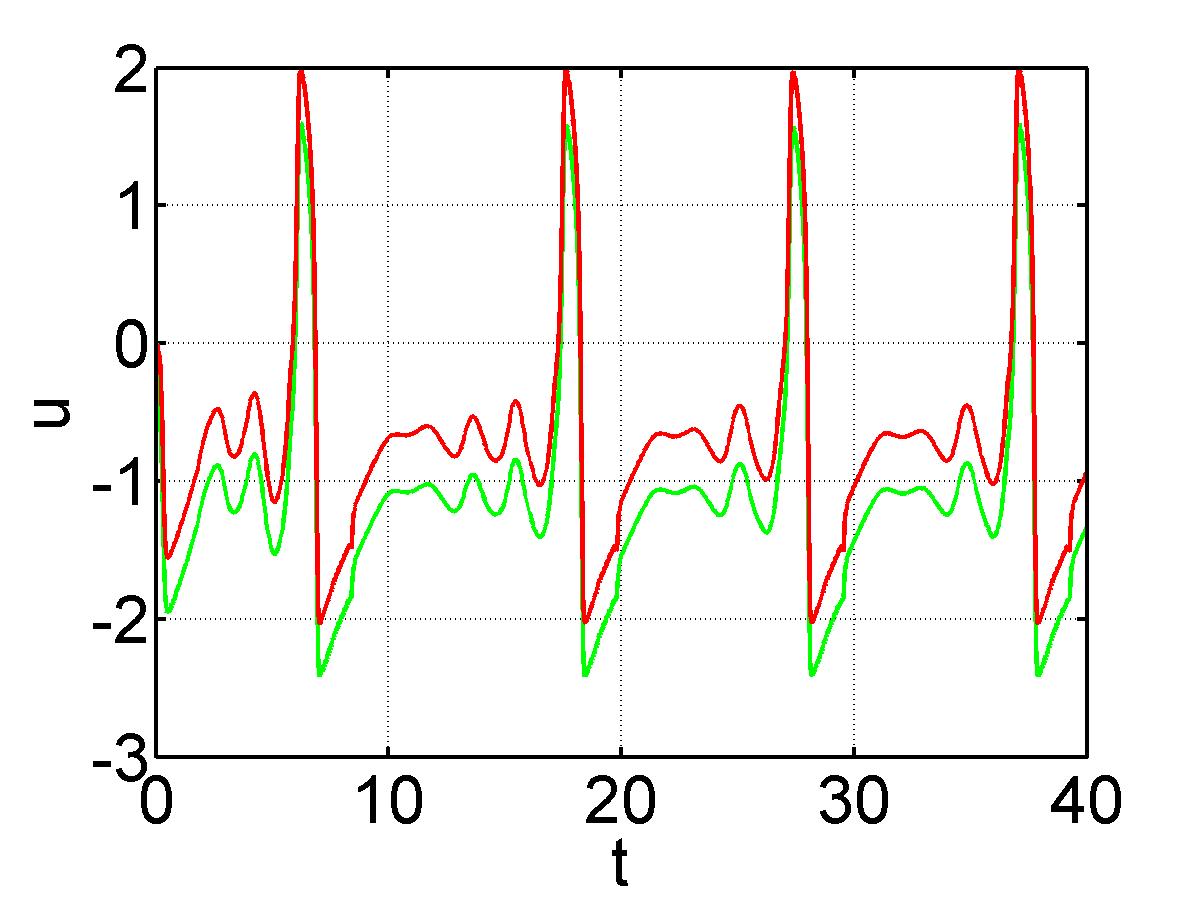} \\ (a)}
\end{minipage}
\begin{minipage}[h]{0.24\linewidth}
\center{\includegraphics[width=1\linewidth]{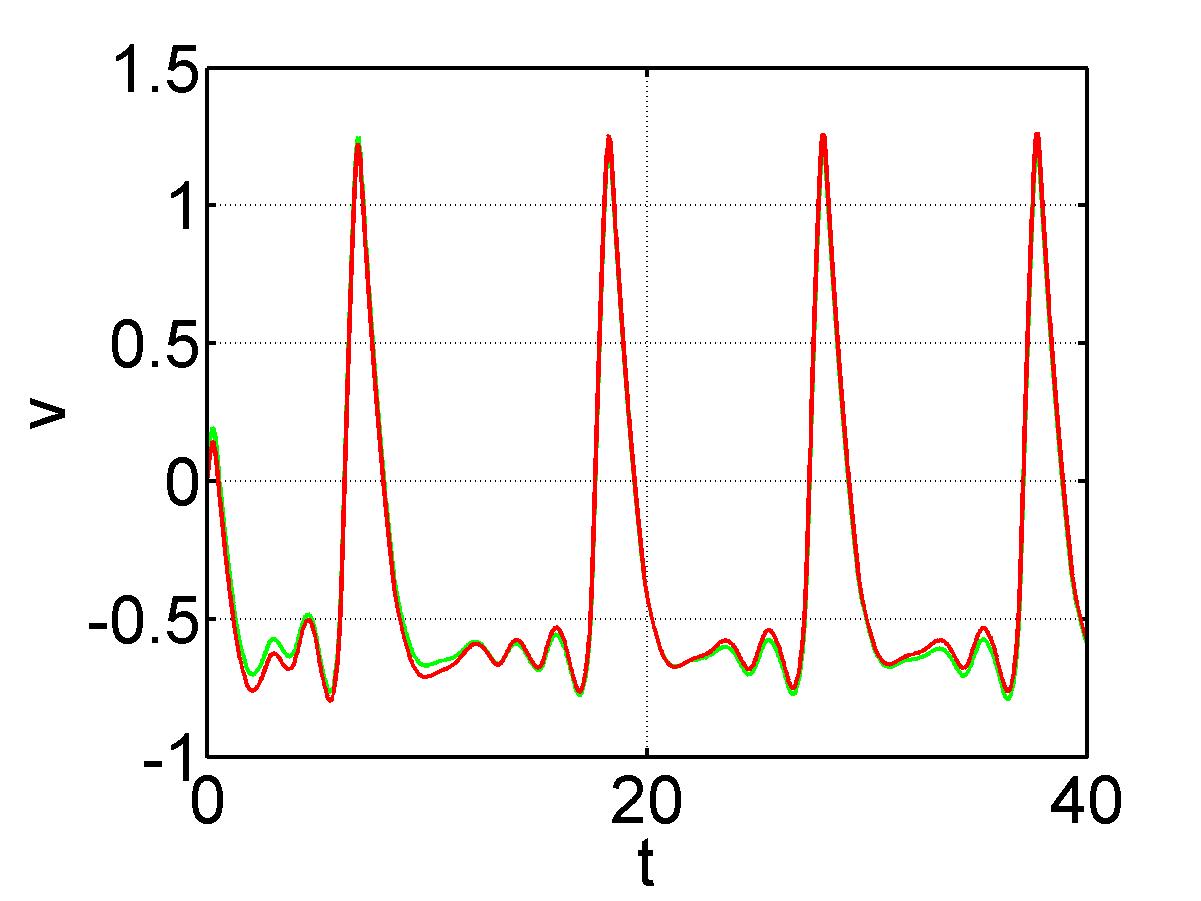} \\ (b)}
\end{minipage}
\vfill
\begin{minipage}[h]{0.24\linewidth}
\center{\includegraphics[width=1\linewidth]{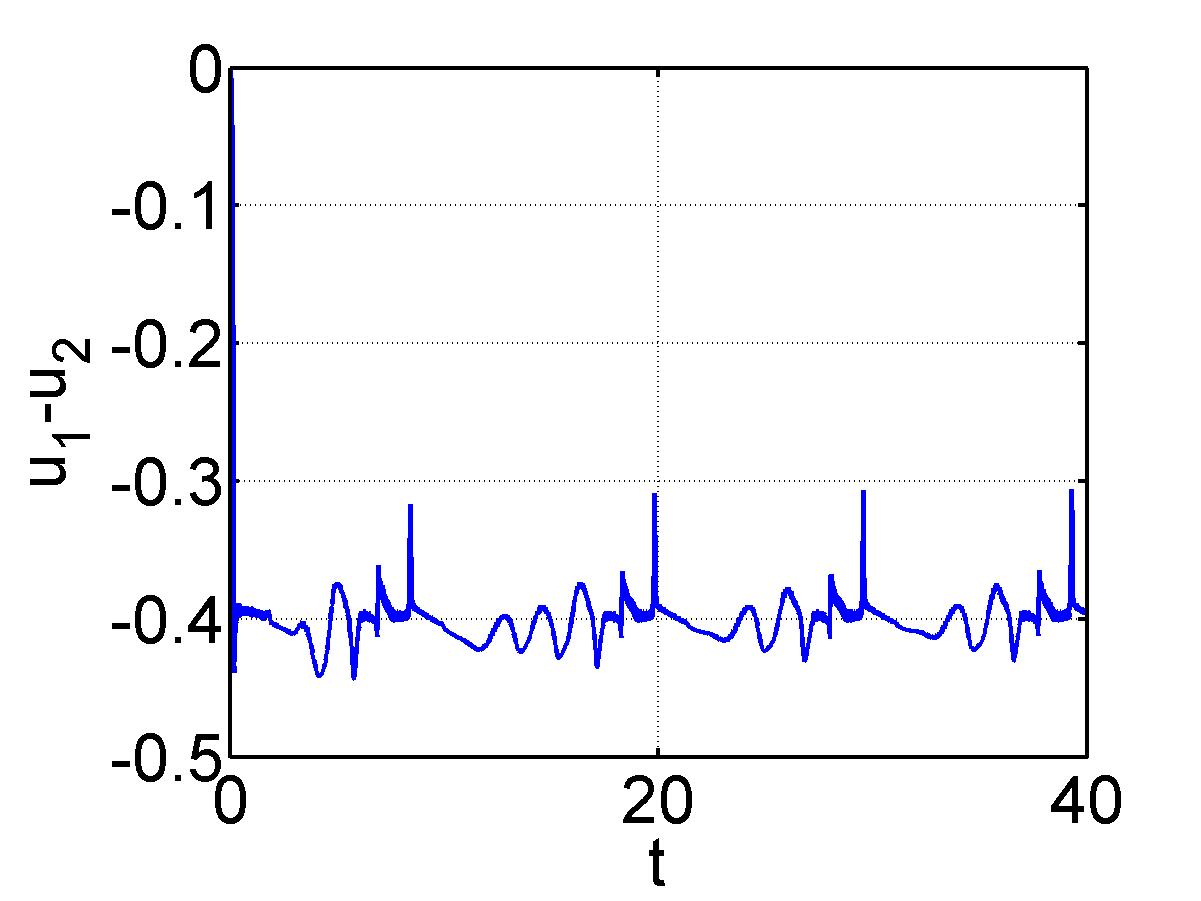} \\ (c)}
\end{minipage}
\begin{minipage}[h]{0.24\linewidth}
\center{\includegraphics[width=1\linewidth]{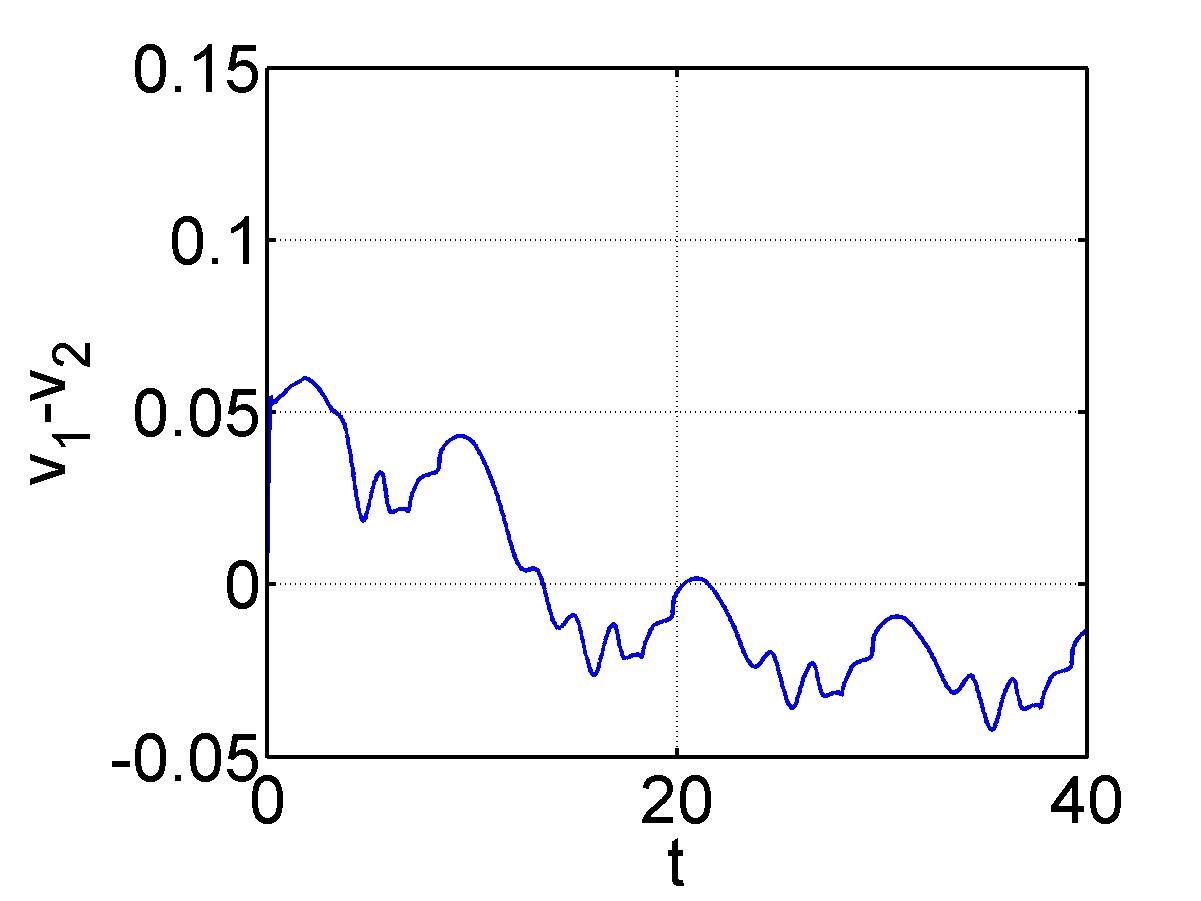} \\ (d)}
\end{minipage}
\caption{
Adaptive control of two coupled FitzHugh-Nagumo systems
(Eq.~\eqref{ad1}) where the coupling strength $C_i$ of each node is
adapted  separately according to Eq.~\eqref{ad2}. 
 Gray (green) 
  line marks node one, black (red)  line marks node two.
(a) and (b): time series of the activator
  and the inhibitor, respectively; (c) and (d): differences $u_1-u_2$
  and $v_1-v_2$ between the activator and the inhibitor values, respectively. Parameters: $\gamma=2$, $C_i^0=0$, $i=1,2$. Other
parameters and initial conditions as in Fig~\ref{fig1}.}
\label{figad1}
\end{figure}

The results of the adaptation according to Eq.~\eqref{ad2}
are shown in Fig.~\ref{figad1}. The two systems reach the desired
synchronized state (see the time series of the activators in
Fig.~\ref{figad1}(a) and the difference between their values in Fig.~\ref{figad1}(c)). Moreover, the difference between inhibitor values is close to zero (see the time series of the inhibitors in
Fig.~\ref{figad1}(b) and the difference between their values in Fig.~\ref{figad1}(d)). Thus, the
control is successful.

\section{Adaptive Synchronization in ring networks}\label{sec:adapt-synchr-ring}
\begin{figure}
\flushleft
\begin{minipage}[h]{0.24\linewidth}
\center{\includegraphics[width=1\linewidth]{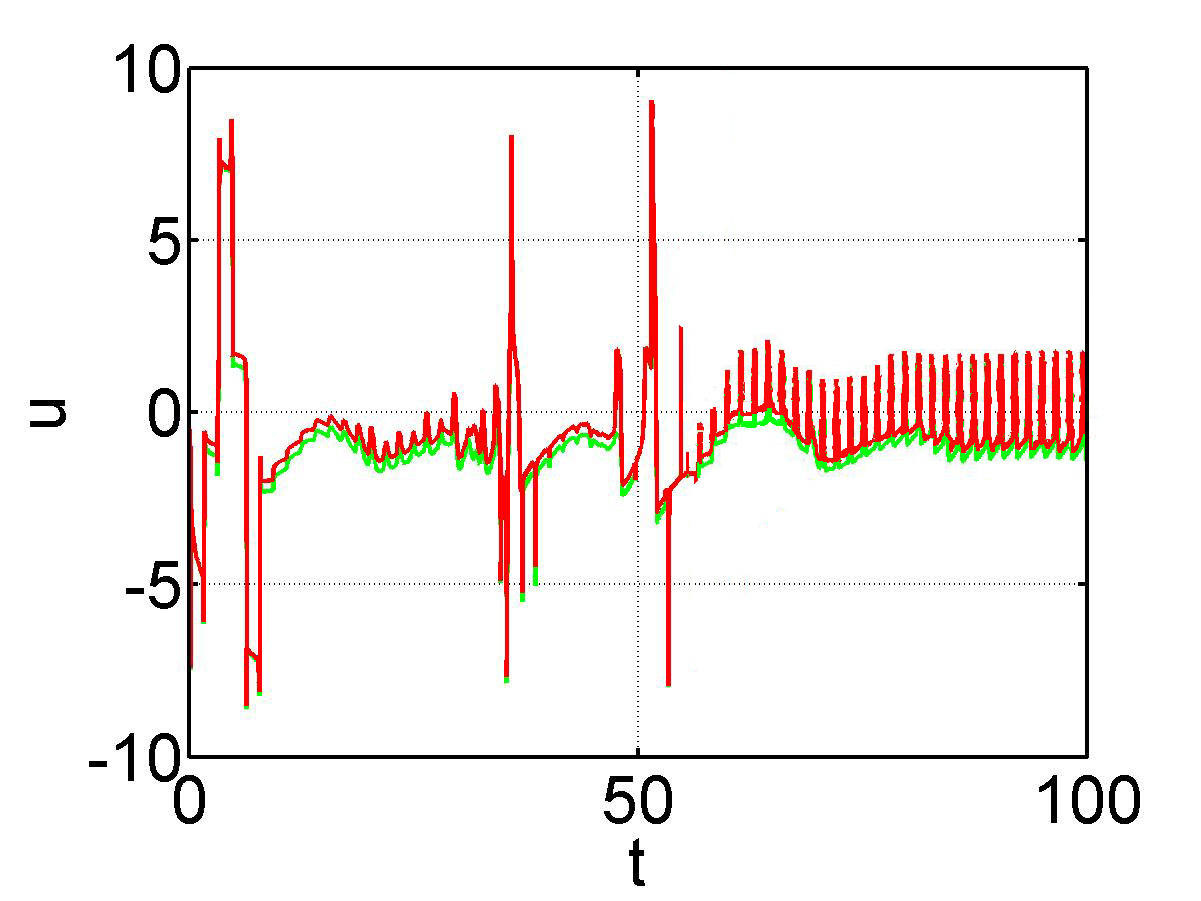} \\ (a)}
\end{minipage}
\begin{minipage}[h]{0.24\linewidth}
\center{\includegraphics[width=1\linewidth]{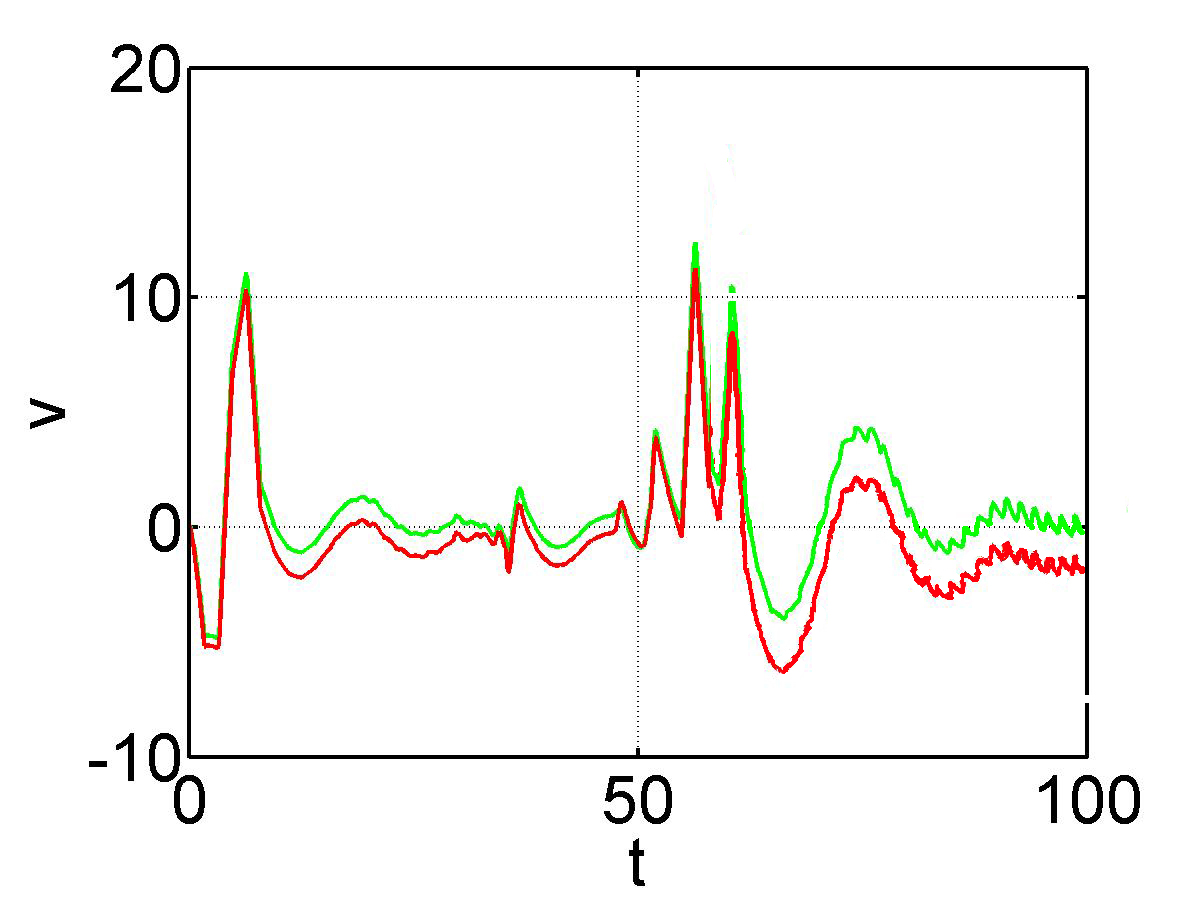} \\ (b)}
\end{minipage}
\vfill
\begin{minipage}[h]{0.24\linewidth}
\center{\includegraphics[width=1\linewidth]{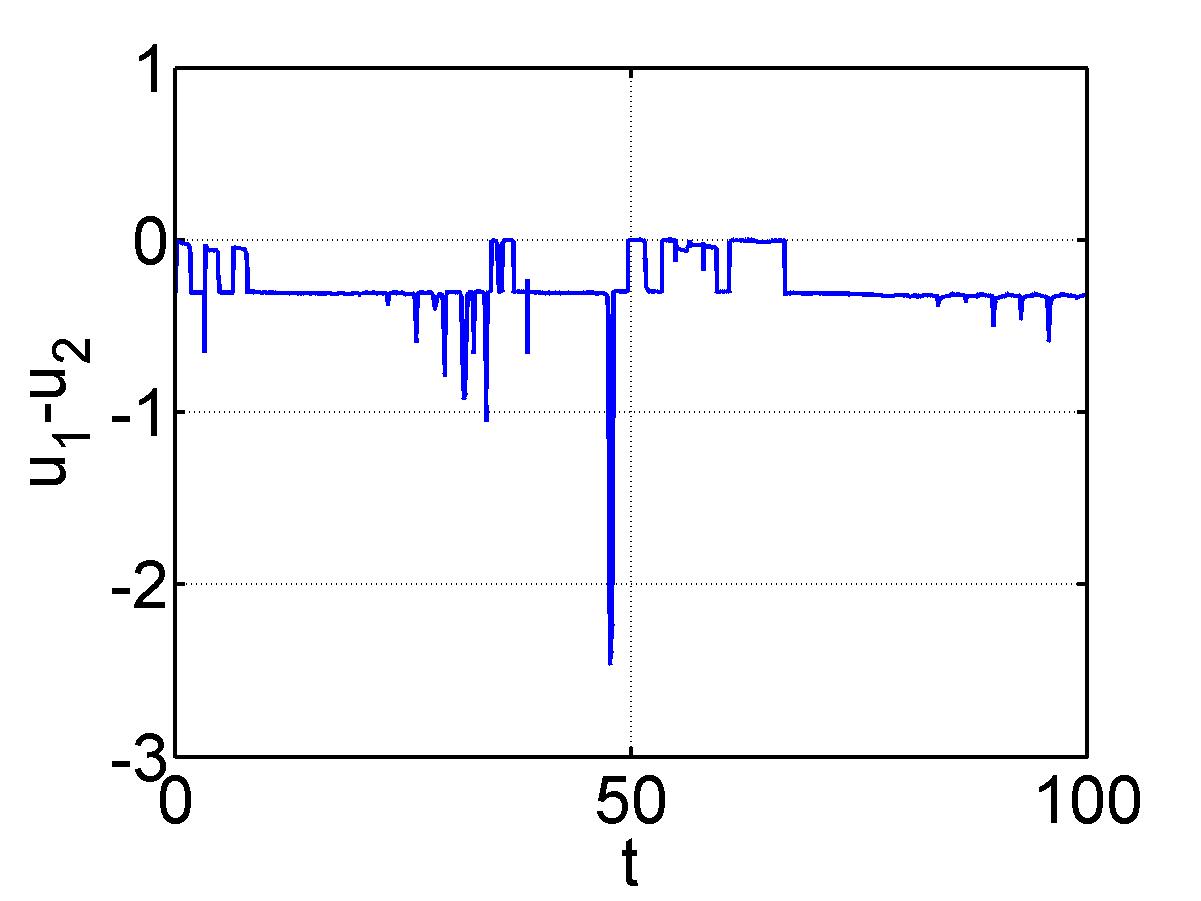} \\ (c)}
\end{minipage}
\begin{minipage}[h]{0.24\linewidth}
\center{\includegraphics[width=1\linewidth]{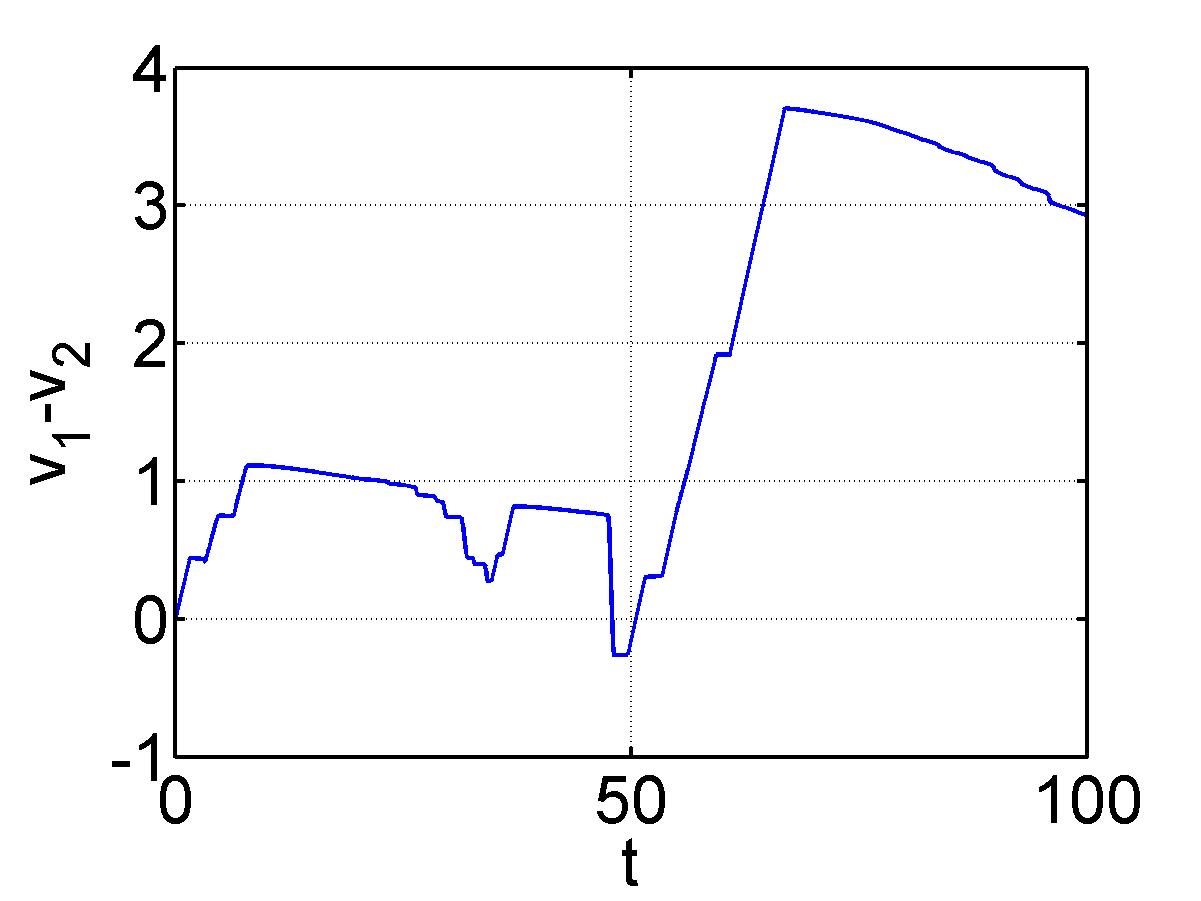} \\ (d)}
\end{minipage}
\vfill
\begin{minipage}[h]{0.24\linewidth}
\center{\includegraphics[width=1\linewidth]{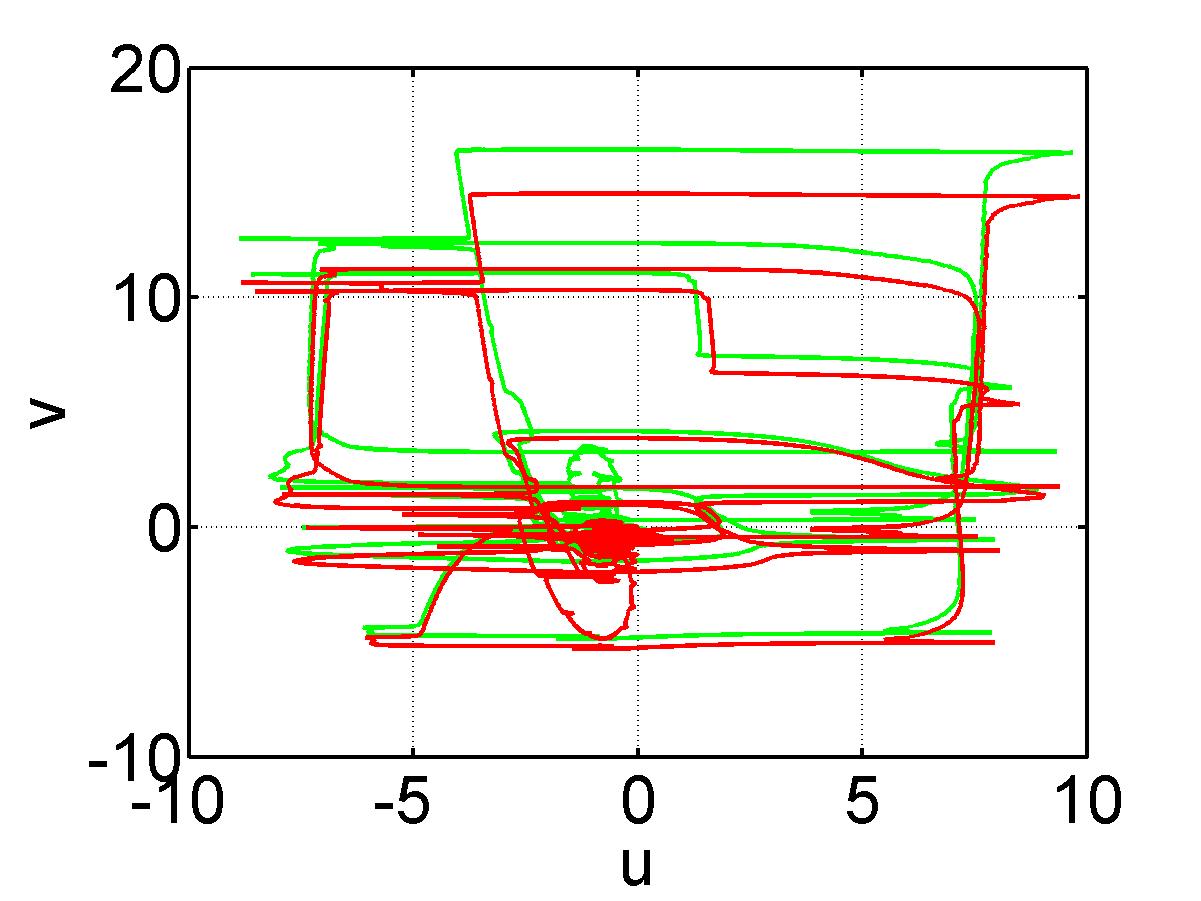} \\ (e)}
\end{minipage}
\begin{minipage}[h]{0.24\linewidth}
\center{\includegraphics[width=1\linewidth]{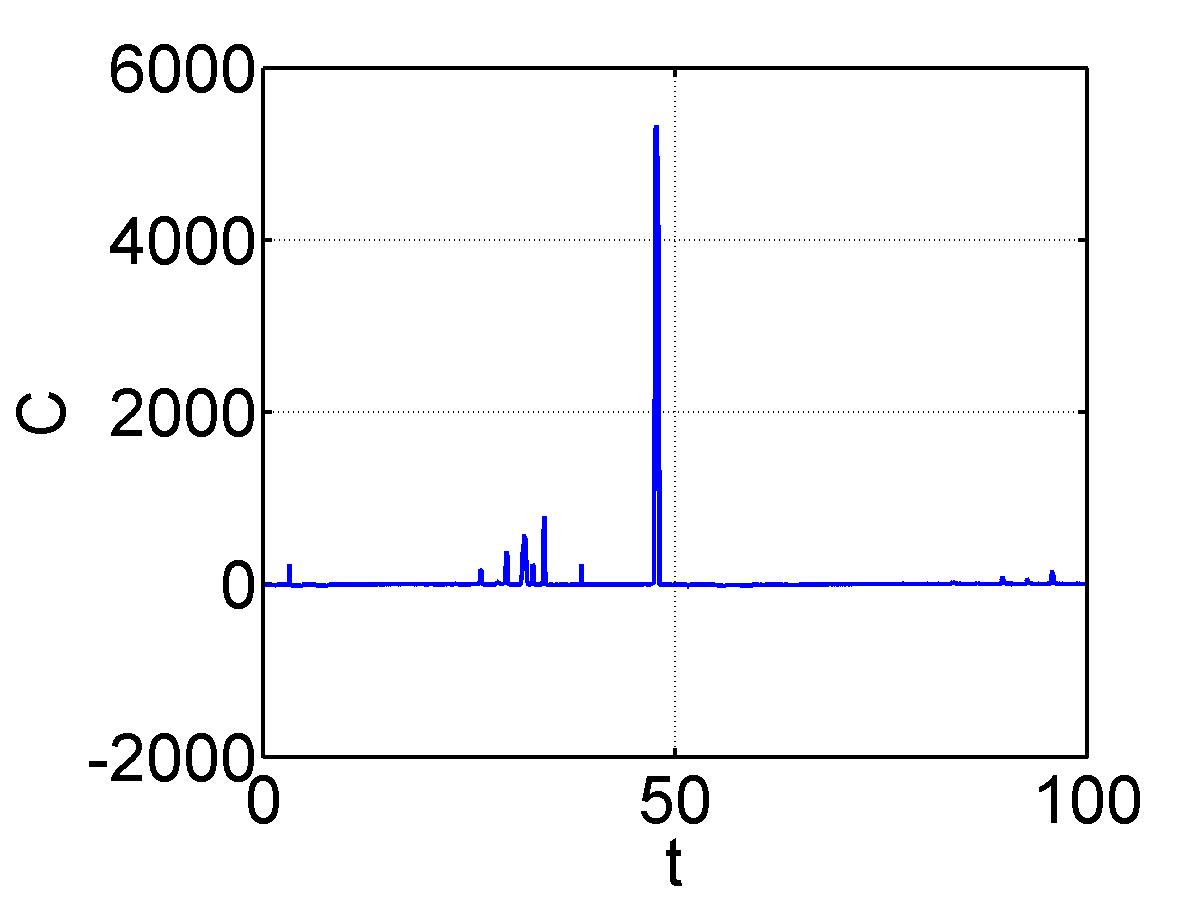} \\ (f)}
\end{minipage}
\vfill
\begin{minipage}[h]{0.5\linewidth}
\center{\includegraphics[width=1\linewidth]{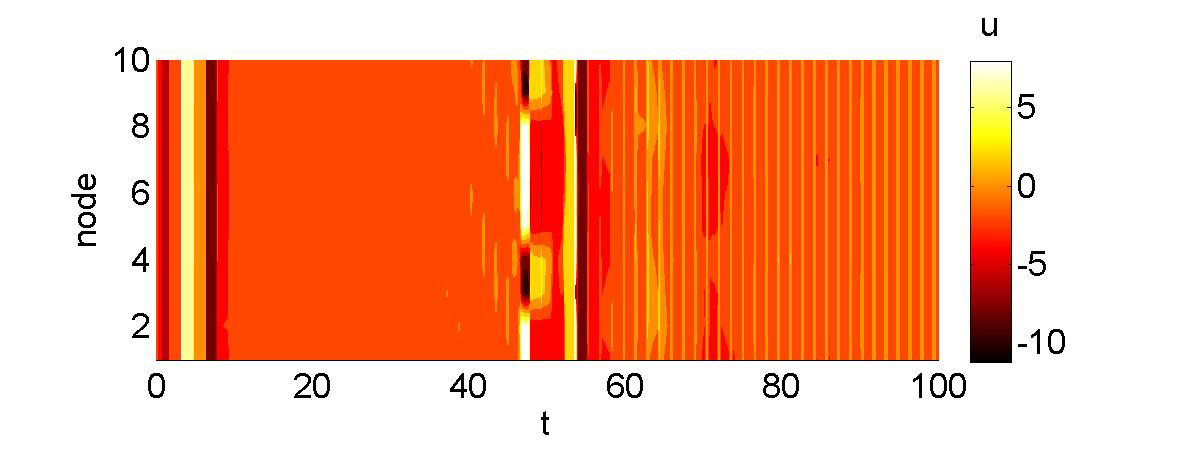} \\ (g)}
\end{minipage}
\vfill
\begin{minipage}[h]{0.5\linewidth}
\center{\includegraphics[width=1\linewidth]{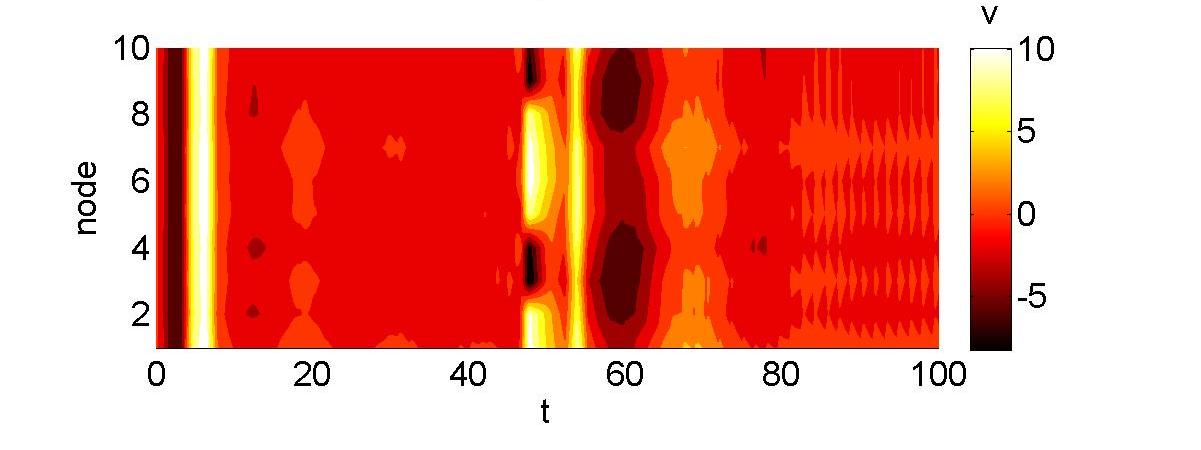} \\ (h)}
\end{minipage}
\caption{Adaptive control of synchronization of a ring of ten FitzHugh-Nagumo
  systems   (according to Eq.~\eqref{m} with coupling matrix \eqref{f12}).
 Gray (green) 
  line marks node one, black (red) line marks node two. (a) and (b): time series of the activator
  and the inhibitor, respectively; (c) and (d): differences between the activator and the inhibitor values, respectively; (e): phase space, (f): time series of
  the coupling strength adapted according to Eq.~\eqref{f13}; (g) and (h): time series of the activators
  and the inhibitors of all nodes, respectively.
 Parameters: $N=10$, $\epsilon=0.1$, $\tau=1.5$, $\gamma=100$,
 $C_0=0$. Threshold parameters $a_i$ are chosen randomly from
 $[0.8,1.1]$, while the highest threshold, here $a_1$,  equals $1.1$
 and the lowest threshold, here $a_2$, equals $0.8$. Initial conditions: $u_i(t)=0$, $v_i(t)=0$, $i=1,\dots,N$,
 for $t\in[-\tau,0]$.}
\label{fig4}
\end{figure}

\begin{figure}
\flushleft
\begin{minipage}[h]{0.5\linewidth}
\center{\includegraphics[width=1\linewidth]{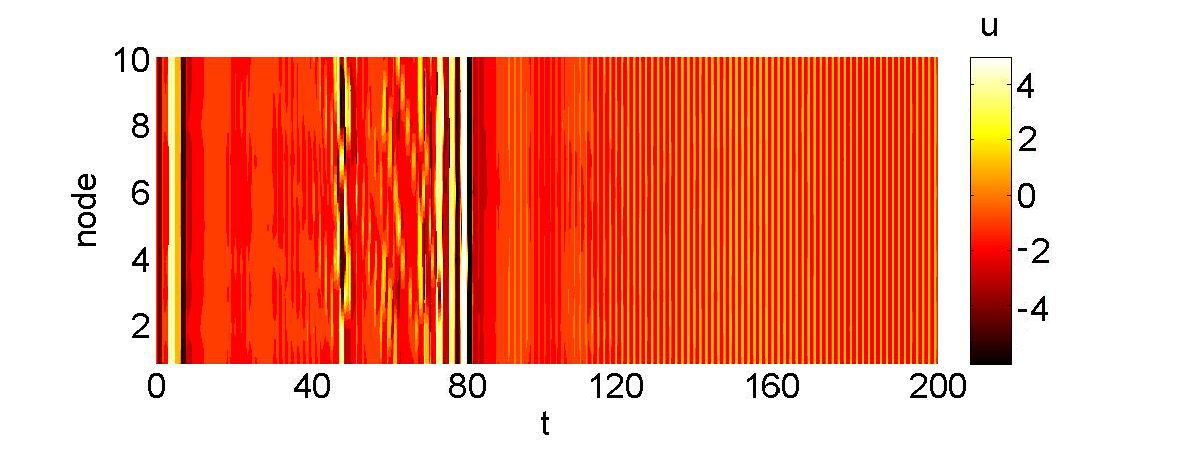} \\ (a)}
\end{minipage}
\vfill
\begin{minipage}[h]{0.5\linewidth}
\center{\includegraphics[width=1\linewidth]{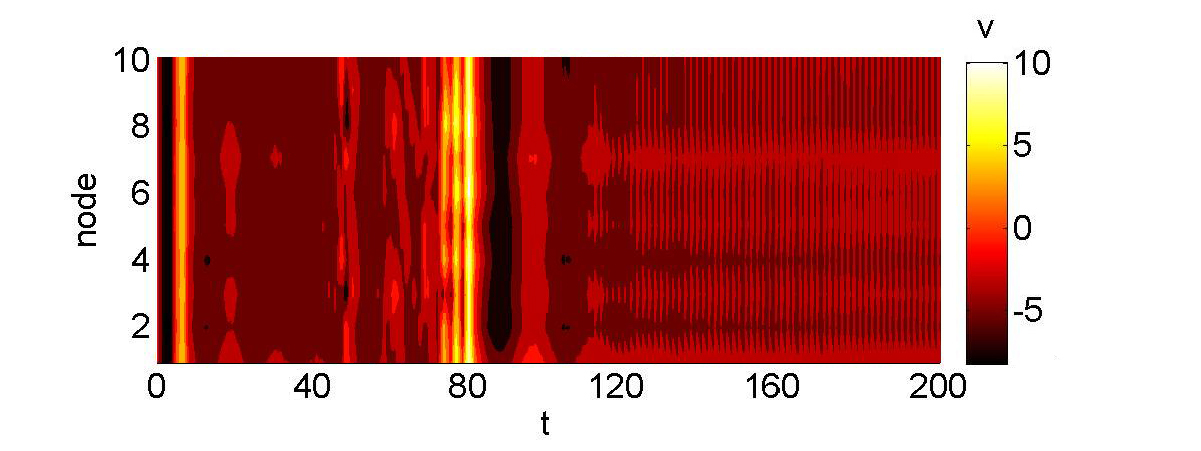} \\ (b)}
\end{minipage}
\caption{As in Fig.~\ref{fig4} but with a bounded coupling strength
  $|C|\le5$. (a) and (b): time series of the activators
  and the inhibitors of all nodes, respectively.}
\label{fig5}
\end{figure}
\begin{figure}
\flushleft
\begin{minipage}[h]{0.5\linewidth}
\center{\includegraphics[width=1\linewidth]{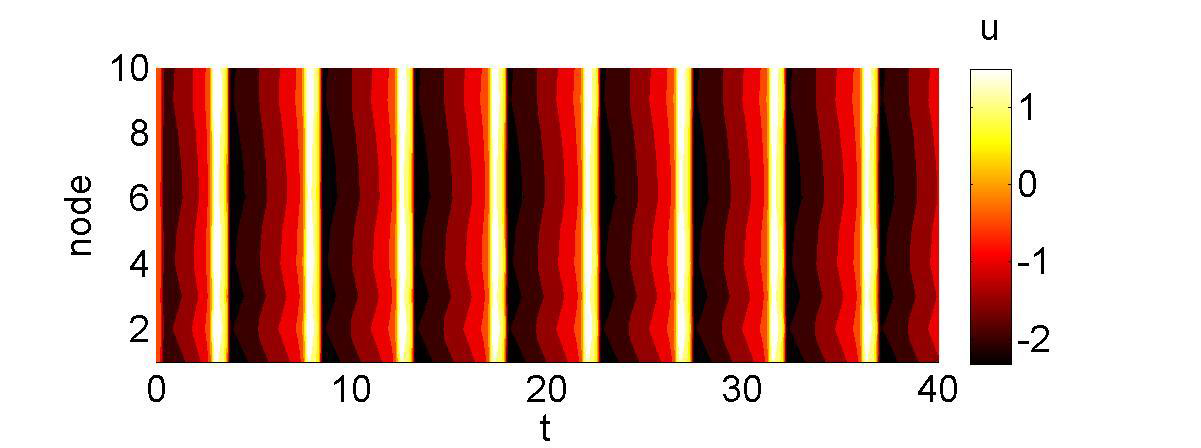} \\ (a)}
\end{minipage}
\vfill
\begin{minipage}[h]{0.5\linewidth}
\center{\includegraphics[width=1\linewidth]{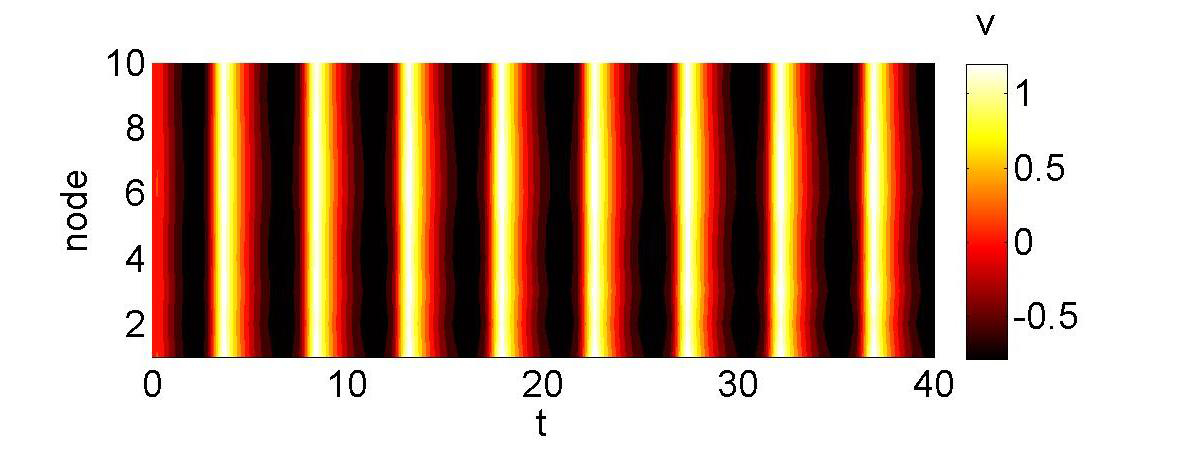} \\ (b)}
\end{minipage}
\caption{Adaptive control of synchronization of ten FitzHugh-Nagumo
  systems where the coupling strength of each node is adapted
  separately according to Eq.~\eqref{f14}. (a) and (b): time series of the activator
  and the inhibitor of all nodes, respectively. Parameters: $C_0^i=0$, $i=1,\ldots,
  N$, $\gamma=1$. Other parameters and initial conditions as in Fig. \ref{fig4}.
}
\label{fig6}
\end{figure}

We now want to apply our method to larger networks. To this end, we
consider a ring network of $N$ nodes where the
coupling matrix  $\mathbf{G}$ has the following form

\begin{equation}\label{f12}
\mathbf{G} = \begin{pmatrix}
0 & 1 & 0 &\cdots & 0 \\
0 & 0 & 1 &\cdots & 0 \\        
\vdots & \vdots & \ddots &\ddots & \vdots \\
0 & 0 & \ddots &\ddots & 1 \\
1 & 0 & 0 &\cdots & 0
\end{pmatrix}.
\end{equation}

Our approach aims to synchronize two nodes of the ring; in this case
the other nodes will follow these two nodes  and synchronize as well. The same idea has been
used in the control of wave motion in a chain of pendulums
\cite{FRA08c}. However, there is a limitation to this approach: The
nodes with the highest threshold and the lowest one have to be
neighbors. 

Let us assume that $k$ is the node with the highest threshold and $l$ is
the one with the lowest threshold, i.e,
\begin{equation}\notag
a_{k}=\max\limits_{i=1,\dots,N}a_i, \qquad a_{l}=\min\limits_{i=1,\dots,N}a_i,
\end{equation}
and that  $a_k$ and $a_l$ are neighbors, i.e.,  $k=(l+1)\bmod N$ or
$k=(l-1) \bmod N$.

We now use the adaptation law~\eqref{f11} 
to synchronize these two
nodes. With nodes $k$ and $l$ instead of nodes  $1$ and $2$, Eq.~\eqref{f11}
reads 
\begin{equation}\label{f13}
C(t) = C_0+\frac{\gamma}{\epsilon}\left[u_k(t)-u_l(t)+a_k-a_l\right]\left[u_k(t)-u_l(t)\right],
\end{equation}
where $\gamma$ is the gain and $C_0$ is an initial value of the coupling
strength. As before, we aim to achieve the control goal \eqref{f2}.
Similar to the case of two coupled FHN systems, the control~\eqref{f13}
ensures synchronization in a nearly synchronized state characterized by 
 \begin{subequations}
 \begin{eqnarray}\label{f8}
 u_i(t)-u_j(t)&\approx& -a_i+a_j, \\
 v_i(t)-v_j(t) &\approx& c_i,\label{f8b}
 \end{eqnarray}
 \end{subequations}
for $t\ge t^*$, where $c_i$ is constant and $i,j=1,\ldots, N$. 

Figure~\ref{fig4} presents the results of a simulation of ten
FitzHugh-Nagumo systems coupled in a ring where the adaption law
\eqref{f13} is applied with $a_1=a_k$ being the node with the highest
threshold, and $a_2=a_l$ being the node with the lowest threshold.  In
Fig.~\ref{fig4}(a) and~\ref{fig4}(c), it is shown that
activators $u_1$ and $u_2$ synchronize.
 Figure~\ref{fig4}(e) shows the phase space of the two nodes. 
In Fig.~\ref{fig4}(g) it can be seen that not only  node one and two
are synchronized but that all nodes follow these two nodes and synchronize after a transient
time. Thus, the control goal is achieved.  

Figure~\ref{fig4}(f) shows that there are several jumps in the control
$C$ during the
transient time and that $C$ temporarily reaches  quite high values. To avoid this, we can limit the coupling strength
$C$. Figure~\ref{fig5} presents the results for a bounded coupling
strength, i.e., $|C|\le 5$. The control is successful; however,  the
transient time  is prolonged  compared to the unbounded control shown in Fig.~\ref{fig4}(g).

So far we considered the case that the nodes with the highest and
lowest coupling strength are neighbors. If this is not fulfilled, the
adaptation of the overall coupling strength does not yield
synchronization. However, if  we control the coupling strength of each
node separately the control goal can be  reached.
The ring network is then described by 
\begin{equation}\label{fhn2}
\begin{aligned}
\epsilon \dot{u_i}&=u_i-\frac{u_i^3}{3}-v_i+C_i(t)[u_{(i+1) \bmod N}(t-\tau)-u_i(t)], \\
\dot{v_i}&=u_i+a_i,\quad i=1,\dots,N,
\end{aligned}
\end{equation}
where $C_i(t)$ describes the strength of the coupling to node~$i$.
As in the case of two nodes, from Eq.~\eqref{f5} with $\mathbf{g}=(C_1,\ldots,C_N)$, goal function 
\begin{align}\label{Qi}
Q(\mathbf{x}(t),t)=&\frac{1}{2}\sum_{i=1}^N\left[u_i(t)-u_{(i+1) \bmod N}(t)+a_i-a_{(i+1) \bmod N}\right]^2,
\end{align}
and
\begin{eqnarray}
\psi(\mathbf{x},\mathbf{g},t)&=&\gamma\nabla_\mathbf{g}\omega(\mathbf{x},\mathbf{g},t),
\end{eqnarray}
we derive the following adaption law
\begin{align}\label{f14}
C_i(t) =&C_i^0+\frac{\gamma}{\epsilon}\left[u_i(t)-u_{(i+1) \bmod N}(t-\tau)\right]\\
&\times\big[2u_i(t)-u_{(i-1) \bmod N}(t)-u_{(i+1) \bmod N}(t)+2a_i
-a_{(i-1) \bmod N}(t)-a_{(i+1) \bmod N}(t)\big],\,i=1,\dots,N, \nonumber
\end{align}
 where $\gamma$ is the gain and $C_i^0$ is the
 initial value of $C_i$.
 
   Figure~\ref{fig6} presents the results of a
 simulation of Eq.~\eqref{fhn2} with adaption law \eqref{f14}. The
 control goal is achieved and synchronization takes place. This method
 requires a lower gain than algorithm~\eqref{f13}, and, moreover, it
 ensures synchronization of the inhibitors with a negligible shift
(see the time series of the inhibitors in Fig.~\ref{fig6}(b)).

\section{Conclusion}\label{sec:conclusion}
We have proposed a novel adaptive method for controlling synchrony in
heterogeneous networks. It is well known that networks with
heterogeneous nodes are much less likely to synchronize than networks of
identical nodes. Furthermore, synchrony will take place in a state
where the trajectories of the different nodes are not identical but
small deviations can be observed. We have suggested a goal function to
characterize this type of synchrony. Based on this goal function and
the speed-gradient (SG) method, we have derived an adaptive controller which
tunes the overall coupling strength such that synchrony is stable
despite the node heterogeneities.

We have demonstrated our method on networks of FitzHugh-Nagumo systems, a
neural model which is considered to be generic for excitable
systems close to a Hopf bifurcation. Before applying the adaptive
control,  we have studied the simple motif of two delay-coupled,
heterogeneous nodes and have given analytic conditions for the
occurrence of the Hopf
bifurcations. We have then applied  our adaptive method and discussed its
dependencies on the node and control parameters. It has been shown that 
 our method enables synchronization even if the node parameters are chosen such
diverse that  one of the systems would exhibit self-sustained oscillations without coupling, while
the other one would remain in a stable equilibrium point, i.e., one of the uncoupled systems
is above, and the other is below the Hopf bifurcation. Furthermore, we have
generalized our method to larger networks and applied it to ring
networks. As an alternative and complement to adapting the overall
coupling strength, we have suggested adapting the coupling strength of each
node separately. This allows for adaptively controlling the network in situations where
the node with the highest threshold is not a direct neighbor of the
node with the lowest threshold, in contrast to the restriction imposed by tuning only the
overall coupling strength.  

Here, we have considered static delays. In principle, it would be
possible to extend the method to time varying delays.
Time varying delay in neural networks has been considered in Refs.~\cite{WAN08d, TAN08a, LI14}. In general networks time varying delay has been
studied in Refs.~\cite{YU08a,GJU14}.  However, to our knowledge  no
adaptive control methods have been applied in networks with time
varying delay and heterogeneous nodes.

Given the paradigmatic nature of the FitzHugh-Nagumo system, we expect
our method to be applicable in a wide range of excitable
systems. Furthermore, the application of the 
SG method to the control of networks with heterogeneous nodes suggests that other
adaptive controllers  that are  based on the SG method (see, for example,
Refs.~\cite{SEL12,SCH12,GUZ13,LEH14}) 
are also robust towards heterogeneities.  

\nonumsection{Acknowledgments}

This work is supported by the German-Russian Interdisciplinary
Science Center (G-RISC) funded by the German Federal Foreign Office
via the German Academic Exchange Service (DAAD). J.L. and E.S.
acknowledge support by Deutsche Forschungsgemeinschaft (DFG) in the
framework of SFB 910. S.P. and A.F. acknowledge that Sec.~\ref{sec:line-stab-fixed} was performed in IPME RAS, supported solely by RSF (Grant No. 14-29-00142).


\end{document}